\documentclass[useAMS,usenatbib]{mn2e}
\usepackage{graphicx}
\usepackage{graphics}
\usepackage{rotating}
\usepackage{amsmath}
\usepackage{amssymb}
\usepackage{lscape}
\usepackage{times}

\title[UCHII and HCHII]{Ultra- and Hyper-compact HII regions at 20 GHz}
\author[T. Murphy et al.]{Tara Murphy$^{1,2}$\thanks{E-mail: tara@physics.usyd.edu.au (TM)},
  Martin Cohen$^{3}$,
  Ronald D. Ekers$^{4}$,
  Anne J. Green$^{1}$, 
  Robin M. Wark$^{4}$, \newauthor
  Vanessa Moss$^{1}$\\
$^{1}$Sydney Institute for Astronomy, School of Physics, The University of Sydney, NSW, 2006, Australia \\
$^{2}$School of Information Technologies, The University of Sydney, NSW, 2006, Australia \\
$^{3}$Radio Astronomy Laboratory, University of California, Berkeley, CA94720, USA \\
$^{4}$Australia Telescope National Facility, CSIRO, P. O. Box 76, Epping, NSW, 1710, Australia }

\newcommand{\changes}[1]{{#1}}

\begin{document}

\date{Accepted 0000 August 12. Received 0000 August 12; in original form 0000 August 12}

\pagerange{\pageref{firstpage}--\pageref{lastpage}} \pubyear{2009}

\maketitle

\label{firstpage}

\begin{abstract}
We present radio and infrared observations of 4 hyper-compact HII regions and 4 
ultra-compact HII regions in the southern Galactic plane. These objects
were selected from a blind survey for UCHII regions using data from two new 
radio surveys of the southern sky; the Australia Telescope 20 GHz survey (AT20G) and the 
2nd epoch Molonglo Galactic Plane Survey (MGPS-2) at 843 MHz.
To our knowledge, this is the first blind radio survey for hyper- and ultra-compact
HII regions.

We have followed up these sources with the Australia Telescope Compact Array 
to obtain H$70\alpha$ recombination line measurements, higher resolution
images at 20~GHz and flux density measurements at 30, 40 and 95~GHz.
From this we have determined sizes and recombination line temperatures as well 
as modeling the spectral energy distributions to determine emission measures.
We have classified the sources as hyper-compact or ultra-compact on the basis of their
physical parameters, in comparison with benchmark parameters from the literature.

Several of these bright, compact sources are potential calibrators for the 
Low Frequency Instrument (30$-$70 GHz) and the 100-GHz channel of the High 
Frequency Instrument of the Planck satellite mission. They may also be useful as 
calibrators for the Australia Telescope Compact Array, which lacks good 
non-variable primary flux calibrators at higher frequencies and in the Galactic plane region.    
Our spectral energy distributions allow the flux densities within the Planck bands to
be determined, although our high frequency observations show that several sources have 
excess emission at 95~GHz (3~mm) that can not be explained by current models.

\end{abstract}

\begin{keywords}
(ISM:) HII regions -- radio continuum (ISM) -- infrared (ISM) -- surveys
\end{keywords}

\section{Introduction}
The process of massive star formation is thought to proceed via collapse and accretion 
from a dense molecular cloud. According to this scenario, a hot molecular gas core 
evolves to produce a massive star within a dense cocoon of gas and dust, most readily 
detected at this early stage by infrared and sub-mm emission. As the Lyman continuum 
from the fledgling star begins to ionize its environs, free-free radio emission is 
detected and the phase designated as an ultra-compact HII region (UCHII) begins. 
\citet{wood89a} first defined this phase observationally and determined that these 
objects were distinct from more evolved HII regions, which are typically optically 
thin at radio wavelengths. UCHII regions are very small ($<0.1$ pc), dense 
($>10^4$ cm$^{-3}$)  ionized regions of gas that surround the youngest and most massive 
O and B stars. The lifetime of an UCHII is estimated to be less than 10$^5$ years 
\citep{comeron96}.

While UCHII regions were thought to be the intermediate stage between a massive 
star being formed in a hot molecular core and a fully formed compact HII region 
of ionised gas, it was discovered by \citet{gaume95} that there is an even earlier 
phase just after the star is newly formed,  but presumably before it becomes an 
UCHII region. In this phase the object is called a hyper-compact HII region (HCHII) 
and is characterised by more extreme parameters, namely electron densities 
($n_e$) $>10^{6} $ cm$^{-3}$, emission measures (EMs) of $>10^{8}$ pc cm$^{-6}$ and 
physical size $<0.05$ pc.  These objects typically have broadened radio recombination 
lines, when compared with the expected line width due to thermal and turbulent 
broadening found in more evolved HII regions. 

Results from \citet{sewilo04}, \citet{kurtz05}, and \citet{gaume95} show that 
the HCHII recombination lines range in width from about 40 to 250~km~s$^{-1}$. 
These broad lines could be explained by bulk motions of gas (infall, outflow or 
rotation), pressure broadening or a blending of components.  \citet{sewilo04} 
studied seven massive star-forming regions and found eight components with broad 
recombination lines, potentially indicating HCHII regions. In some sources a 
double-peaked profile may indicate infalling gas as an ionized accretion flow 
\citep{keto03}. A halo of diffuse gas, optically thin at 3.6~cm, surrounding
the HCHII candidates was a surprise as was the presence of multiple massive stars 
at the same early evolutionary phase, given their expected brief lifetimes. 

Such a precise definition of an HCHII region is somewhat artificial and the 
question of whether there is merely a continuum in evolutionary phases or whether 
HCHII represent a distinct class of objects remains unresolved. For example, 
\citet{kurtz05} speculates that HCHII regions are signposts for individual stars, 
whereas the objects classified as UCHII regions often contain clusters. 

The currently identified HCHII regions are heavily optically obscured by a thick 
cocoon of dust and are very rare, principally for two reasons. Firstly, there is 
a strong selection bias in the early surveys searching for young HII regions. These 
surveys, mostly around $\lambda$ 6~cm, were optimised for objects with electron 
densities about $10^{4} $ cm$^{-3}$. At higher densities an object remains optically 
thick to a much higher frequency and will below the detection limit \citep{kurtz05}.
The second constraint is that the expected lifetime of this phase is likely to be brief 
\citep{comeron96} and potentially difficult to capture.

In this early phase of massive stellar evolution, no optical and minimal near-infrared 
light emerges from the core because of the enormous amount of surrounding dust.  
Radiative transfer maintains a large gradient in dust temperature from the core 
to the outer surface of the dust shell and the region emits brightly from the far- to 
mid-infrared (FIR, MIR) regime.  If there is a continuum of evolutionary stages, 
then as the star ages the HCHII expands, becoming first an UCHII and then a compact 
HII region, defined as having a size of $<0.5\,\rm{pc}$. Subsequently, the FIR$-$MIR 
emission fades and eventually the dust becomes optically thin at visible wavelengths 
and the birth process is complete.

The stellar radiation from the HCHII and UCHII phases is absorbed by the cocoon of  
molecular gas and dust and re-emitted in the far-infrared. As a result, they are 
amongst the brightest point sources in the IRAS 100~$\mu$m catalogue.
At radio wavelengths the surrounding molecular gas and dust is transparent, 
and so UCHII regions appear as bright unresolved continuum sources. It is not clear 
at what stage the HCHII regions become bright radio sources, as they may be optically 
thick at lower frequencies. The continuum emission is often associated with OH and 
H$_2$O masers, which indicates that the UCHII regions are affected by stellar winds, 
bow shocks, masers and other dynamical processes in star formation regions. 
Some HCHII regions are also associated with the masers, particularly 
H$_2$O masers which are produced by dense gas. For example G75.78$+$0.34 
\citep{hofner96} and NGC\,7538 (see summary in \citet{sewilo04}).

Although UCHII were initially defined on the basis of being unresolved 
in radio continuum observations, high resolution  ($<1\arcsec$) radio observations 
have revealed a variety of morphologies. 
These have been classified into 5 general types by \citet{wood89a}: 
Spherical (or unresolved), Cometary (parabolic), Core-halo, Shell and Irregular 
(or multiply peaked). The apparent shape of an UCHII is affected by the optical 
depth of the gas at the particular observing frequency. When gas is optically 
thick, only the surface of the UCHII region is visible, and the internal structure 
is hidden.  UCHII usually have an inverted (or rising) 
spectral index at radio frequencies and are much brighter in the FIR/MIR regime.

The observational morphologies for HCHII regions are less well known. From the 
literature, we have collated a working definition of HCHII regions as a stage in 
development earlier than UCHII (see Section \ref{s_criteria}).
The difficulty in determining morphologies is that the sources are largely unresolved. 
The fit to a simple one component model and its limitations are discussed later in 
this paper.

The objects presented here are drawn from a complete sample of 46 sources identified 
as part of a blind survey for UCHII regions (Murphy et al., in preparation). 
The initial selection of sources in the blind survey was done on the basis of their 
rising spectral index between 843~MHz and 20~GHz. 
We have used this subset of eight objects to investigate the process of classifying and 
characterising the sources as UCHII and HCHII regions. In addition to being 
scientifically interesting, these objects are also likely to be useful calibrators for 
the Planck satellite mission, as well as for high frequency observations with the 
Australia Telescope Compact Array.

In Section \ref{s_radio} we describe our observations and sample selection procedure, 
incorporating data from the  Molonglo Observatory Synthesis Telescope (MOST) and the 
Australia Telescope Compact Array (ATCA). For this paper the sub-sample were studied in 
detail to validate the process of characterizing the complete sample in a future project.  
Follow-up observations with the ATCA were undertaken to refine the positions and flux 
densities of these eight objects. Section \ref{s_analysis} gives our radio frequency 
results and analysis of radio recombination line data. Section \ref{s_seds} 
presents the spectral energy distributions of the eight regions and the models that best 
fit our multi-frequency results. In Section \ref{s_criteria}, the 
benchmark definitions of HCHII and UCHII regions are generated from a synthesis of parameters 
collected from the literature and our classifications are discussed.
Section \ref{s_mir} presents a comparison of the mid-infrared (MIR) and radio observations.

\section{Radio Frequency Observations}\label{s_radio}

Our sample was selected using observations from two new panoramic surveys of 
the Southern sky; the second epoch Molonglo Galactic Plane Survey \citep[MGPS-2;][]{murphy07}
\citep{murphy07} which covers the sky south of Declination $-30\degr$ at 843~MHz and 
the Australia Telescope 20~GHz survey  \citep{massardi08,murphy10}. The two surveys 
are well matched in resolution ($45\arcsec$ and $30\arcsec$ respectively), and are 
described in more detail below.
We carried out a blind search for UCHII, selecting a sample of 46 sources 
on the basis of being bright ($f_{20GHz} \ge 200\,\rm{mJy}$), compact, and having a rising 
(inverted) spectral index between 843~MHz and 20~GHz ($\alpha^{20}_{0.843} > 0.1$).

Once a list of candidates was produced, follow-up observations of 20~GHz continuum  
and the H$70\alpha$ radio recombination line were made to provide accurate 
positions and flux densities and to eliminate extragalactic sources. The database 
SIMBAD was used to remove known stars. Eight of the remaining Galactic 
sources were selected for subsequent observation and analysis to test the potential 
of this method for identifying objects in the earliest phases of HII region evolution. 
These sources provide the link between massive star formation from hot molecular cores 
and the evolved HII regions which figure prominently in optical, infrared and radio 
images of the Galaxy. The eight sources selected were identified as the 
best candidates for further investigation as they were isolated sources with a MIR 
image relatively unencumbered by diffuse Galactic emission.  

\subsection{The Australia Telescope 20 GHz survey}\label{s_at20g}
The Australia Telescope 20~GHz survey (AT20G) is a blind 
20~GHz survey of the whole southern sky ($\delta \le 0\degr$) including the 
Galactic plane. Follow-up observations were carried out at 5, 8 and 20~GHz for 
all detections excluding those in the Galactic plane ($|b| > 1.5\degr$).
The survey was carried out between 2004 and 2008, detecting approximately 
5800 sources down to the 40~mJy flux density limit. 

The AT20G was carried out in two phases. An initial blind survey was done in a
special scanning mode of the ATCA. Using custom data reduction and source extraction
software, a list of positions and fluxes for candidate sources brighter than 
$5\sigma$ (about 40~mJy) was produced. This scanning survey is described in 
(Hancock et al, in preparation). 

For the region outside the Galactic plane ($|b| > 1.5\degr$), each candidate 
source was re-observed with the ATCA in standard snapshot mode to confirm the
detection and measure accurate positions, flux densities (at 5, 8 and 20~GHz) and 
polarization information \citep{massardi08,murphy10}.
The results presented here are part of a larger program to investigate 
the Galactic plane at 20~GHz as this region was excluded from the main AT20G 
follow-up survey because of the complexity of the emission in the Plane.

\subsection{The Molonglo Galactic Plane Survey}
The second epoch Molonglo Galactic Plane Survey \citep[MGPS-2;][]{murphy07} is the 
Galactic counterpart to the Sydney University Molonglo Sky Survey 
\citep[SUMSS;][]{bock99,mauch03} and together the surveys cover the whole sky south 
of $\delta=-30\degr$ at a frequency of 843~MHz. They were undertaken over the period 
1997 to 2007 with the Molonglo Observatory Synthesis Telescope 
\citep[MOST;][]{mills81,robertson91}. MGPS-2 has better resolution 
($45\arcsec\times45\arcsec\csc|\delta|$) and sensitivity than any previous panoramic 
radio survey of the southern Galaxy. 

The primary data products are a catalogue of compact sources, with $48\,850$ sources 
above a limiting peak flux of 10 mJy beam$^{-1}$ \citep{murphy07} and a set of mosaic
images. Positions in the catalogue are accurate to $1-2\arcsec$ and flux densities to 
about $5\%$. Images from MGPS-2 
and SUMSS are available online \footnote{http://www.physics.usyd.edu.au/sifa/Main/MGPS}\footnote{http://www.physics.usyd.edu.au/sifa/Main/SUMSS}.

\subsection{Follow-up Radio Observations}
All follow-up observations of the eight sources were carried out with the 
Australia Telescope Compact Array (ATCA) at Narrabri.
The observations are described below and a summary of their technical specifications
is given in Table~\ref{t_obsparms}. The output from each run was processed using 
standard techniques with the {\sc miriad} data reduction package. 
\begin{table*}
\centering
\caption{Summary of observation parameters. 
The three pairs of high frequency continuum measurements were each combined to give a single 
image and flux density for that frequency band. Note that 7~mm data for  G301.1366-00.2248 was
obtained in a different configuration, as described in Section \ref{s_high}.}\label{t_obsparms}
\begin{tabular}{lcccccc}\hline
Frequency (GHz)      & 18.496 \& 19.520 & 18.624     & 18.769  & 32.064 \& 34.112 & 42.944 \& 44.992 & 93.504 \& 95.552 \\
Array configuration  &  1.5C            &   H75      &  H75  &  H214  &  H214 & H214 \\  \hline
Observing mode       & continuum        &  continuum & line      &  continuum & continuum & continuum \\
Bandwidth (MHz)      & 128              & 128        & 32        &   128  &   128  & 128 \\
No. Channels         & 32               & 32         & 128       &   32  &    32  & 32 \\
Primary beam         & 2\farcm7         & 2\farcm7   & 2\farcm5 & 1\farcm5 & 1\farcm1 & 0\farcm5 \\
Synthesised beam     & $1\farcs7\times1\farcs2$ & $34\arcsec\times25\arcsec$ & $34\arcsec\times34\arcsec$  & $6\farcs7\times6\farcs7$ & $5\farcs0\times5\farcs0$ & $2\farcs3\times2\farcs3$ \\
Observing dates      & 2007 Apr 26, 27  & 2006 Sep 16$-$18 & 2006 Sep 16$-$18 & 2008 Jul 10 & 2008 Jul 10 & 2008 Jul 10 \\ 
\hline
\end{tabular}
\end{table*}

\subsubsection{Snapshot imaging}
The AT20G flux densities from the blind survey scanning runs are accurate to 
$\sim20\%$. Two sets of follow-up measurements at 20 GHz were necessary to confirm 
the flux density and spectral index selection criteria of the sources and also to 
provide high resolution images to compare with the infrared data (see Section \ref{s_mir}). 

On 2006 September 18 our sources were observed with the ATCA in the H75 configuration, 
which is a compact hybrid array with baselines ranging from 31~m to 89~m.
The synthesised beam is $34\arcsec\times25\arcsec$ which is well matched to MGPS-2 
(resolution $\sim45\arcsec$) and the US Midcourse Space eXperiment \citep[MSX;][]{Price01} 
with a resolution of $\sim20\arcsec$. We simultaneously observed at 18.624 GHz (continuum 
mode with 128 MHz bandwidth split into 32 channels) and 18.769 GHz (spectral line mode with 
128 channels over 32 MHz bandwidth, see Section \ref{s_recom}).
With this configuration, each source was observed in 3 cuts, with a total integration 
time of 4 minutes per source. As expected, all of the sources in our sample were 
unresolved with this array.

\begin{figure*}
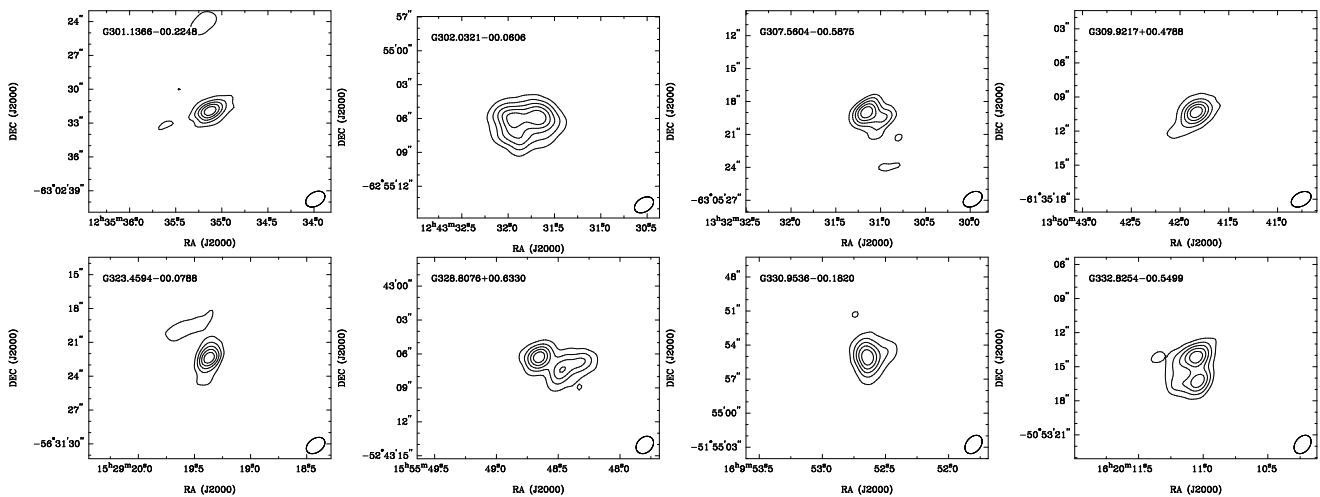

\centering
\includegraphics[height=4.3cm,angle=270]{images/1235-6302.19008.ps}
\includegraphics[height=4.3cm,angle=270]{images/1243-6255.19008.ps}
\includegraphics[height=4.3cm,angle=270]{images/1332-6305.19008.ps}
\includegraphics[height=4.3cm,angle=270]{images/1350-6135.19008.ps}
\includegraphics[height=4.3cm,angle=270]{images/1529-5631.19008.ps}
\includegraphics[height=4.3cm,angle=270]{images/1555-5243.19008.ps}
\includegraphics[height=4.3cm,angle=270]{images/1609-5154.19008.ps}
\includegraphics[height=4.3cm,angle=270]{images/1620-5053.19008.ps}
\caption{High resolution images of our UCHII and HCHII regions.
The 18.496 and 19.520 GHz data have been combined to make a single 19.008 GHz
image. The contours show 6 levels, equally spaced between $5\sigma$ and the image
peak.}
\label{f_highres}
\end{figure*}

The second set of snapshot observations was made on 2007 April 26 and 27 with 
the ATCA in the 1.5C configuration (with baselines of 77~m to 4500~m). 
Each source was observed for $8-12$ short periods spread over 12 hours, with a total 
integration time of 1 hour per source. This gave us the UV coverage necessary to
produce high resolution ($\sim1.5\arcsec$ FWHM) images, which are shown in 
Fig.~\ref{f_highres}.

\subsubsection{H$70\alpha$ radio recombination lines}\label{s_recom}
\begin{figure*}
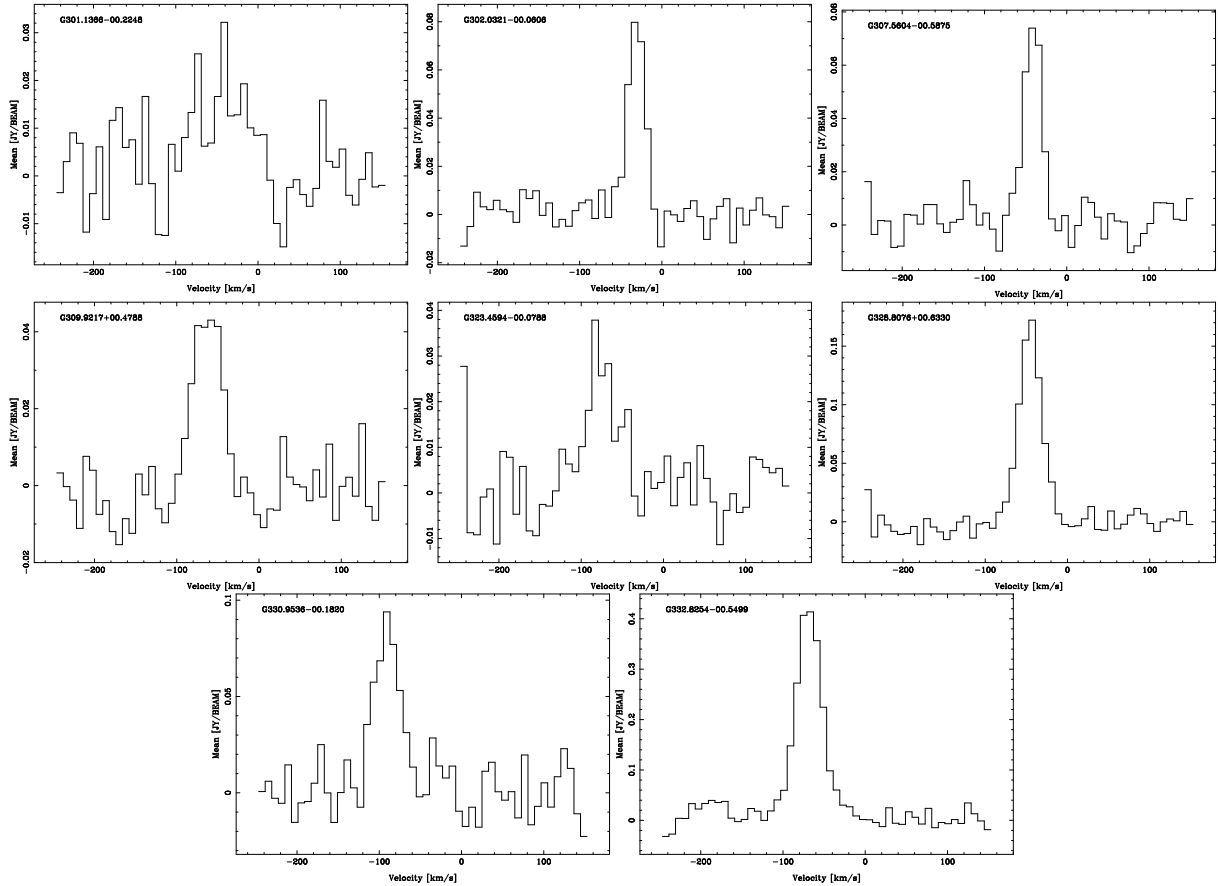

\centering
\includegraphics[height=5.3cm,angle=270]{images/G301.1366-00.2248.rrl.ps}
\includegraphics[height=5.3cm,angle=270]{images/G302.0321-00.0606.rrl.ps}
\includegraphics[height=5.3cm,angle=270]{images/G307.5604-00.5875.rrl.ps}
\includegraphics[height=5.3cm,angle=270]{images/G309.9217+00.4788.rrl.ps}
\includegraphics[height=5.3cm,angle=270]{images/G323.4594-00.0788.rrl.ps}
\includegraphics[height=5.3cm,angle=270]{images/G328.8076+00.6330.rrl.ps}
\includegraphics[height=5.3cm,angle=270]{images/G330.9536-00.1820.rrl.ps}
\includegraphics[height=5.3cm,angle=270]{images/G332.8254-00.5499.rrl.ps}
\caption{H$70\alpha$ radio recombination line spectra for our UCHII and HCHII regions
showing the mean flux density in a $2\times2$ pixel region centred on the source
position.}
\label{f_rrl}
\end{figure*}

Hydrogen recombination lines are a customary diagnostic for HII regions. In the 2006 
September 18 snapshot observing run with the H75 hybrid array, we used the second
frequency band to observe the H$70\alpha$ radio recombination line (rest frequency 
18.769~GHz). 
The resulting velocity resolution was 250~kHz (4 km s$^{-1}$) with a total velocity 
coverage for each spectrum measurement of $\sim$500 km s$^{-1}$. For the H$70\alpha$ 
transition in a typical HII region, the observed spectral line is predicted to be 
approximately  5\% of the peak flux of the associated continuum source, which was well 
within our sensitivity limits. The eight sources presented here all had a clear H$70\alpha$ detection,
as shown in the spectra plotted in Fig.~\ref{f_rrl}.

\subsubsection{High frequency observations}\label{s_high}
On 2008 July 10 we observed seven of the eight sources (excluding G301.1366-00.2248) at 33~GHz 
(central frequencies 32.064~GHz and 34.112~GHz) and 44~GHz (central frequencies 42.944~GHz and 
44.992~GHz) as part of the ongoing ATCA calibrator program (C007). We also observed
seven of the eight sources at 95~GHz (central frequencies 93.504~GHz and 95.552~GHz).

All observations were done in continuum mode with a 128 MHz bandwidth on each of the two central 
frequencies, sub-divided into 32 channels. We used 
the H214 hybrid array, with baselines from 92~m to 247~m. 
\changes{We observed one or two cuts for each object, with a typical total integration time of 
a few minutes. 
This is not enough to do meaningful imaging, but is sufficient to measure the angular size and 
flux density accurately. The smallest scale these observations are sensitive to is
roughly 5--6\arcsec at 33~GHz and 44~GHz and 2\arcsec at 95~GHz.}

The planet Uranus was used as the primary flux calibrator. Calibration at these high 
frequencies can be problematic, which is one of the motivations for searching for new UCHII
candidates as potential flux calibrators for the ATCA.

\changes{The remaining object, G301.1366-00.2248, was observed on 2009 September 01 as part of the
C2050 calibrator program. We obtained data at 33, 44 and 46~GHz using the EW352 array, with
baselines from 31~m to 352~m.}

\subsubsection{Radio properties of the sample}
The observational attributes of the sources are shown in Table~\ref{t_seds}.
These include our measured flux densities in Janskys; the Gaussian-fitted angular 
sizes at 18.624\,GHz, ($\theta$) in arcsec; the recombination line-to-continuum ratios 
($F_L$/$F_C$); $V$, the LSR radial velocity of the recombination line in km~s$^{-1}$;  
and, in the final column, the ratios of spatially integrated 8.0-$\mu$m fluxes to those at 843\,MHz.   
This ratio provides a discriminant between thermal and non-thermal radio emission 
\citep{cohen01}.  MIR/radio continuum ratios have been measured for ensembles of thermal 
emission sources, in particular for HII regions of different morphologies \citep{cohen01} 
and planetary nebulae \citep{cohen07}.  Non-thermal sources, for example galaxies 
and supernova remnants, are about 400 times weaker than a thermal source of equivalent
radio intensity.

The errors in the flux densities are a function of the RMS noise, systematic errors in the 
calibration, and errors associated with our method of measurement. The listed flux 
densities at high frequencies are dominated by systematic calibration uncertainties, which are
hard to estimate. In Table~\ref{t_seds} we have given an overall estimate of the percentage 
error at each frequency.
\begin{table*}
\begin{center}	
\caption{Measured attributes of the sources, including flux densities in Janskys, recombination
line ratios ($F_L/F_C$) and velocities (V) in km~s$^{-1}$, and MIR/radio flux ratios. }\label{t_seds}
{\small
\begin{tabular}{cccrrrrrrrrr}\hline
Name                & $f_{0.843}$    & $f_{18.624}$&  $\theta$('')&  $F_L$/$F_C$& V&$ f_{19.008}$  & $f_{32.064}$ & $f_{42.944}$         & $f_{44.992}$          & $f_{95}$        & MIR/$f_{0.843}$\\
                    & ($\pm5\%$)    & ($\pm10\%$) & & & & ($\pm10\%$) & ($\pm10\%$) & ($\pm10\%$) & ($\pm10\%$) & ($\pm10\%$) \\
\hline
G301.1366$-$00.2248 & 0.012 & 1.03 & 0.7&  0.053& $-20$ & 0.9 &   1.6   & 1.7   & 1.8    &  $-$   & $200\pm30$ \\

G302.0321$-$00.0606 & 0.274 & 0.99 & 4.0& 0.219& $-31$ & 1.0 & 1.0 & 1.0 & 0.9 & 1.2 & $45\pm2$ \\

G307.5604$-$00.5875 & 0.446 & 0.74 & 8.0& 0.224& $-41$ & 0.8 & 0.8 & 0.8 & 0.7 & 0.8 & $40\pm2$ \\

G309.9217$+$00.4788 & 0.028 & 1.02 & 1.0& 0.163& $-63$ & 1.1 & 1.1 & 1.2 & 1.1 & 3.1 & $450\pm 50$ \\

G323.4594$-$00.0788 & 0.094 & 0.98 & 2.0& 0.106& $-75$ &  $-$     & 1.3 & 1.3 & 1.3 & 3.1 & $90\pm10$ \\

G328.8076$+$00.6330 & 0.224 & 1.91 & 4.4& 0.203& $-44$ &  $-$     & 2.1 & 2.1 & 2.0 & 4.2 & $30\pm3$ \\

G330.9536$-$00.1820 & 0.151 & 3.77 &  1.8& 0.102& $-91$ & 5.0 & 6.0 & 6.8 & 6.5 & 21.0 & $16\pm2$ \\

G332.8254$-$00.5499 & 0.669 & 4.76 & 3.0& 0.189& $-69$ & 7.1 & 6.0 & 6.0 & 5.8 & 14.5 & $80\pm10$ \\
\hline
\end{tabular}
}
\end{center}
\end{table*}

\section{Analysis}\label{s_analysis}

Recombination lines can be characterised as Gaussian profiles with a 
peak $F_L$, a central velocity $V$ and a FWHM velocity $\Delta V$. 
The line velocity can be used to estimate a kinematic distance for the object, 
which then determines its linear size and subsequent classification.
The line width is used to calculate the electron temperatures ($T_e$) and any 
broadening becomes a useful discriminant between HCHII and UCHII regions.
The derived parameters calculated in this section are given in Table~\ref{t_models}.
\changes{Also included in Table~\ref{t_models} are extinction values calculated using the
optical depth of silicate features in the IRAS Low Resolution Spectrometer (LRS)
spectra, as discussed in Section~\ref{s_extinct}.}
\begin{table*}\centering
\caption{Summary of quantities derived from pure free-free, constant density model 
fits to observations from 843~MHz$-$45~GHz.  \label{t_models}}
\begin{tabular}{crrrrrrrrcc}\hline     
Name& $\alpha^{20}_{0.843}$& T$_e$& $\Delta$V& Dist.& Size& EM& N$_e$& A$_V$& 95/45~GHz & Class \\
      &                    &  (K) &  (km s$^{-1}$)& (kpc)& (pc) & (pc cm$^{-6}$) & & (mag) & flux ratio & \\ \hline
G301.1366$-$00.2248 & 1.5 & 11600 & 66 &  4.5 & 0.02 & 3.0E+09 & 6.8E+05 & $11\pm1$ & ...& H\\
G302.0321$-$00.0606 & 0.4 & 8000  & 24 &  7.2 & 0.14 & 3.0E+07 & 1.5E+04 & $20\pm2$ & 1.6& U\\
G307.5604$-$00.5875 & 0.2 & 7100  & 27 &  7.7 & 0.30 & 5.0E+06 & 4.1E+03 & $60\pm5$ & ...& U \\
G309.9217$+$00.4788 & 1.2 & 6700  & 40 &  5.5 & 0.03 & 8.0E+08 & 2.3E+05 & $40\pm3$ & 2.0& H \\
G323.4594$-$00.0788 & 0.8 & 8100  & 50 &  4.8 & 0.05 & 1.9E+08 & 6.4E+04 & $18\pm1$ & 2.9& H\\ 
                    &     &       &    &  8.9 & 0.09 &         & 4.7E+04 &          &     & U--H\\ 
G328.8076$+$00.6330 & 0.7 & 6400  & 34 & 11.7 & 0.25 & 6.0E+07 & 1.6E+04 & $50\pm4$ & 2.5& U\\
G330.9536$-$00.1820 & 1.1 & 10400 & 38 &  5.5 & 0.05 & 1.3E+09 & 1.6E+05 & $9\pm1$   & 3.5&  H\\ 
                    &      &       &    &  9.3 & 0.08 &        & 1.2E+05 &          &    &  U--H\\
G332.8254$-$00.5499 & 0.7 & 6500  & 36 &  4.4 & 0.06 & 4.0E+08& 8.0E+04  & ...      &  1.8& U \\
\hline
\end{tabular}
\end{table*}

\subsection{Electron and brightness temperatures}
The continuum and recombination line brightness temperatures $T_C$ and 
$T_L$ can be calculated from the peak flux density using the 
Rayleigh-Jeans approximation
\begin{equation}
T_B = \frac{c^2}{2\nu^2 k_B \Omega} S_\nu
\end{equation}
where $\Omega$ is the solid beam angle. 

We calculated the electron temperature $T_e$ for each object following 
\citep{condon07} (Equation 7C5):
\begin{equation}
T_e \simeq \left[7\times10^3 \nu^{1.1} \frac{1}{1.08 \Delta v}\frac{T_c}{T_L} \right]^{0.87}
\end{equation}
which assumes LTE and that the typical He$+$/H$+$ ion ratio is 
N(He$+$)/N(H$+$)$ \simeq 0.08$. These derived values are shown in Table \ref{t_models}
and are primarily of interest for the SED modelling in the next section.
\changes{Note that since the recombination line width may be affected by other factors
than just the temperature, our values of $T_e$ may be overestimated. However, this 
overestimate can not be large, or our SED's will not fit the observed flux densities 
and sizes. For our data the uncertainties due to complexity of source shapes make $T_e$ 
estimates from the continuum alone unreliable but this does indicate that future higher 
quality and higher resolution imaging observations with flux densities and size measurements 
at a range of frequencies could be used to separate thermal broadening of the recombination 
lines from other effects.}

\subsection{Distance estimates}\label{s_distance}
We calculated kinematic distances using the Galactic rotation curve of 
\citet{mcclure-griffiths07}
\begin{equation}
\Theta(R) = \left(0.186\frac{R}{R_0} + 0.887\right)\Theta_0
\end{equation}
with $\Theta_0 = 220{\rm km s}^{-1}$ and the distance to the Galactic 
centre $R_0 = 8.5{\rm kpc}$. Four of our sources (G302.0321$-$00.0606, 
G307.5604$-$00.5875, G328.8076$+$00.6330, and G332.8254$-$00.5499) are 
included in the \citet{caswell87} study of southern HII regions, and our 
distances are in good agreement.

For those of our sources with a distance ambiguity, we followed \citet{caswell87} 
who resolved a number of such ambiguities by checking for optical counterparts.
The argument is that sources with optical counterparts must be at the near distance
because dust would obscure any object further than about 5~kpc. We took the same 
approach using the highly sensitive SuperCOSMOS H$\alpha$ Survey \citep{Parker05} 
and have attempted to refine the distance to which an object can be detected in 
H$\alpha$ as a function of Galactic longitude.
We also consulted 2MASS K$_s$ images to look for diffuse counterparts. 
The distance of 4.4~kpc for G332.8254$-$00.5499 was considered most likely based on 
a visual identification by \citet{caswell87}.

Frew (priv. comm.) has determined reddenings for about 1000 ESO PNe listed by 
\citet{acker92} and \citet{acker96}.  These diminish around $E(B-V)$\footnote{$E(B-V)$ is the 
{\it reddening} or {\it colour excess} and is a measure of the difference between the
apparent and unobscured colours of an object.} of 3 (i.e. for extinction A$_V \ge 9.3$).
Frew notes that recently discovered PNe are impossible to detect in H$\alpha$ when 
A$_{H\alpha}$ exceeds 7 mag., equivalent to A$_V\sim$ 8.6~mag, consistent with his
estimate from  ESO PNe.  Therefore, PNe can serve as a probe of the interstellar medium's (ISM)
average transparency.  Mean values of extinction near the Galactic plane have been measured
as a function of longitude by \citet{Joshi05} and vary from about 0.2 to 0.6 mag\,kpc$^{-1}$.  
The lowest reddening occurs for longitudes between 190\degr and 290\degr and the highest
toward the Galactic Centre \citep{Arenou92,Chen98,Joshi05}.  For our HII regions, we expect to
detect H$\alpha$ emission to a distance of 12, 6, and 5\,kpc at longitudes 
of 270\degr, 300\degr and 330\degr, respectively.  Non-detection of H$\alpha$ suggests that
distances greater than these are more likely in a given direction.  Using these guidelines
we have resolved most of the distance ambiguities in Table~\ref{t_models}, putting the most
likely distance in the table. 
When we have been unable to decide (e.g. in the absence of 
any optical and even a near-infrared counterpart), two models for that source have been calculated,
corresponding to both distances.

\section{SED models}\label{s_seds}
\begin{figure*}
\includegraphics[width=7cm]{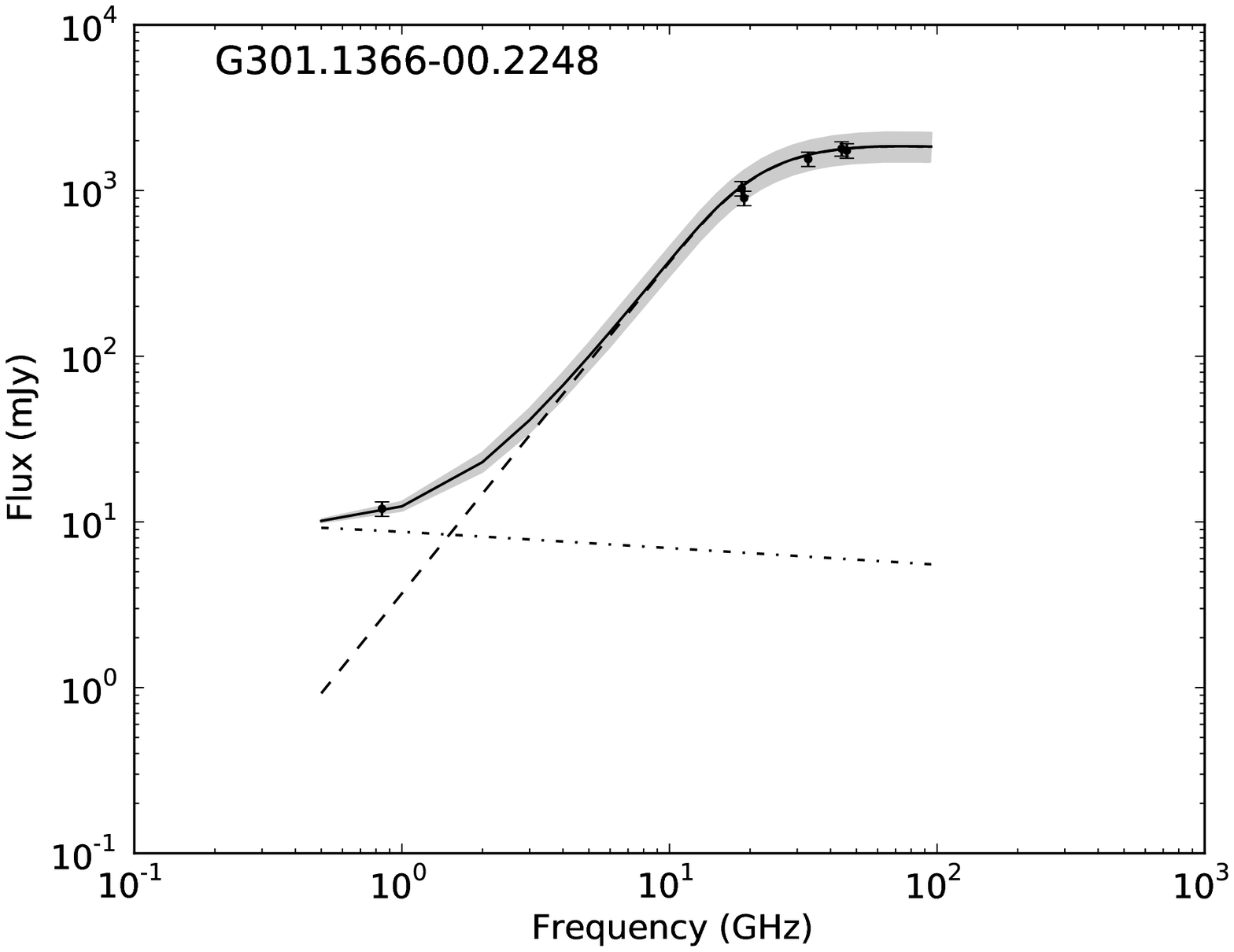}
\includegraphics[width=7cm]{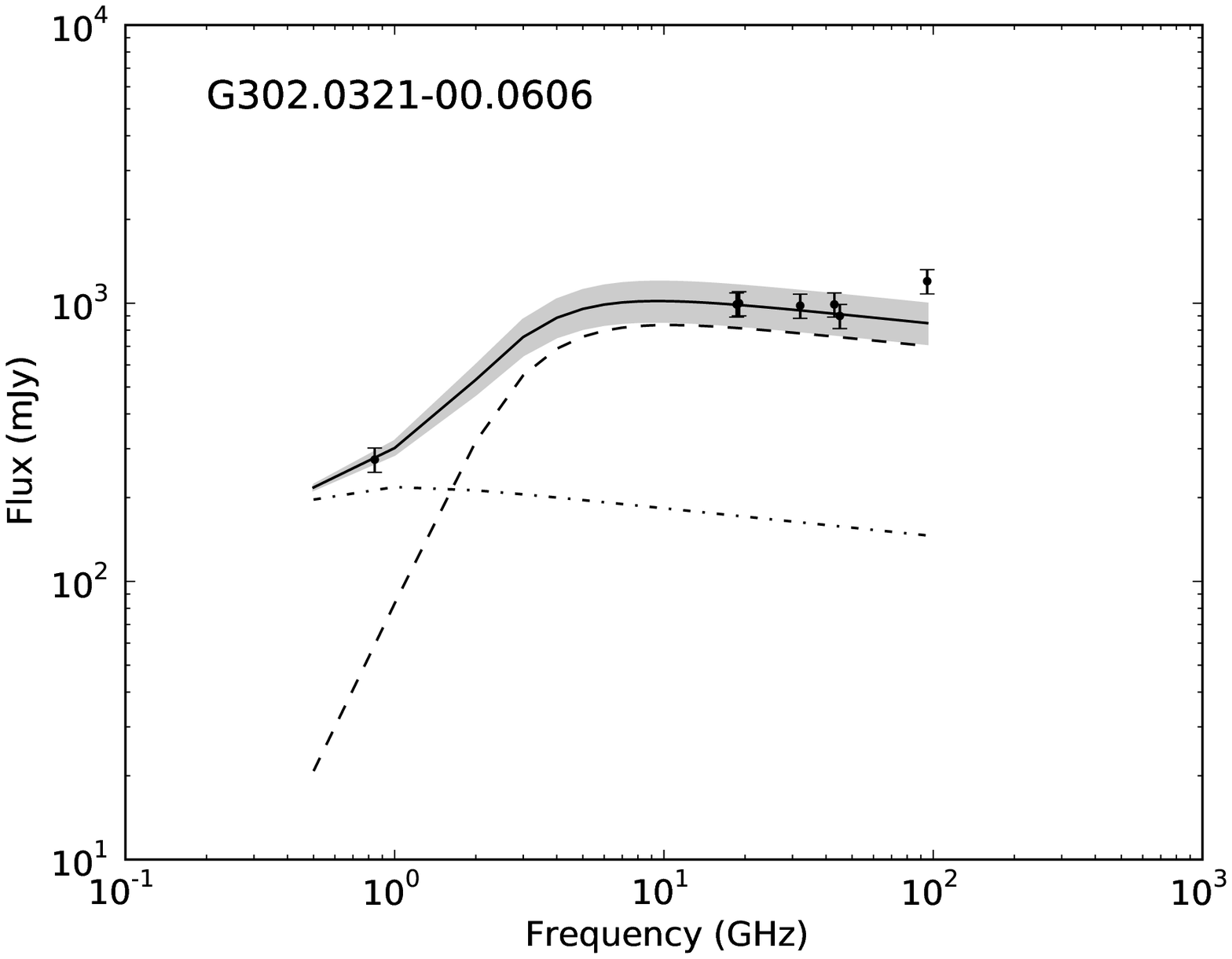}
\includegraphics[width=7cm]{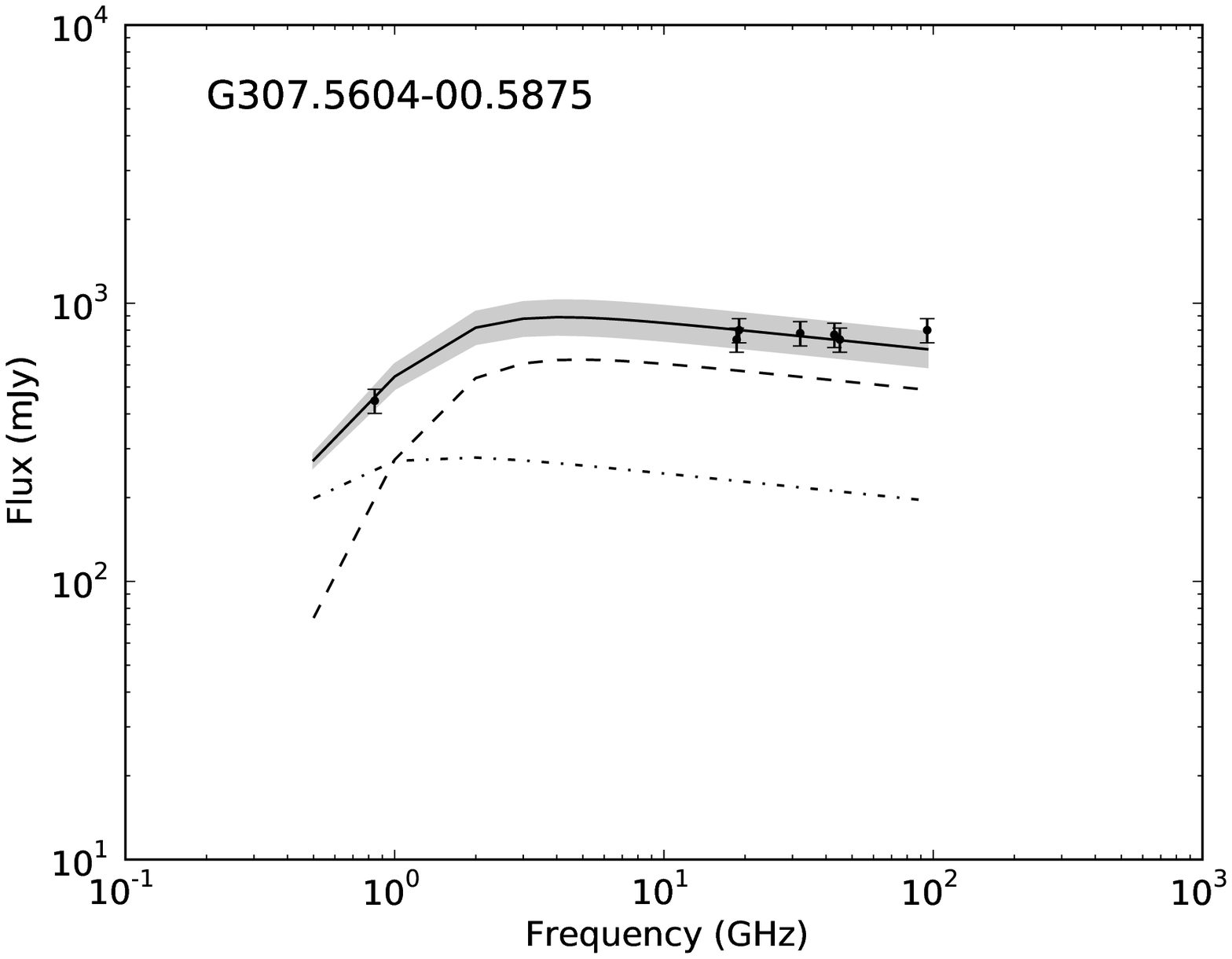}
\includegraphics[width=7cm]{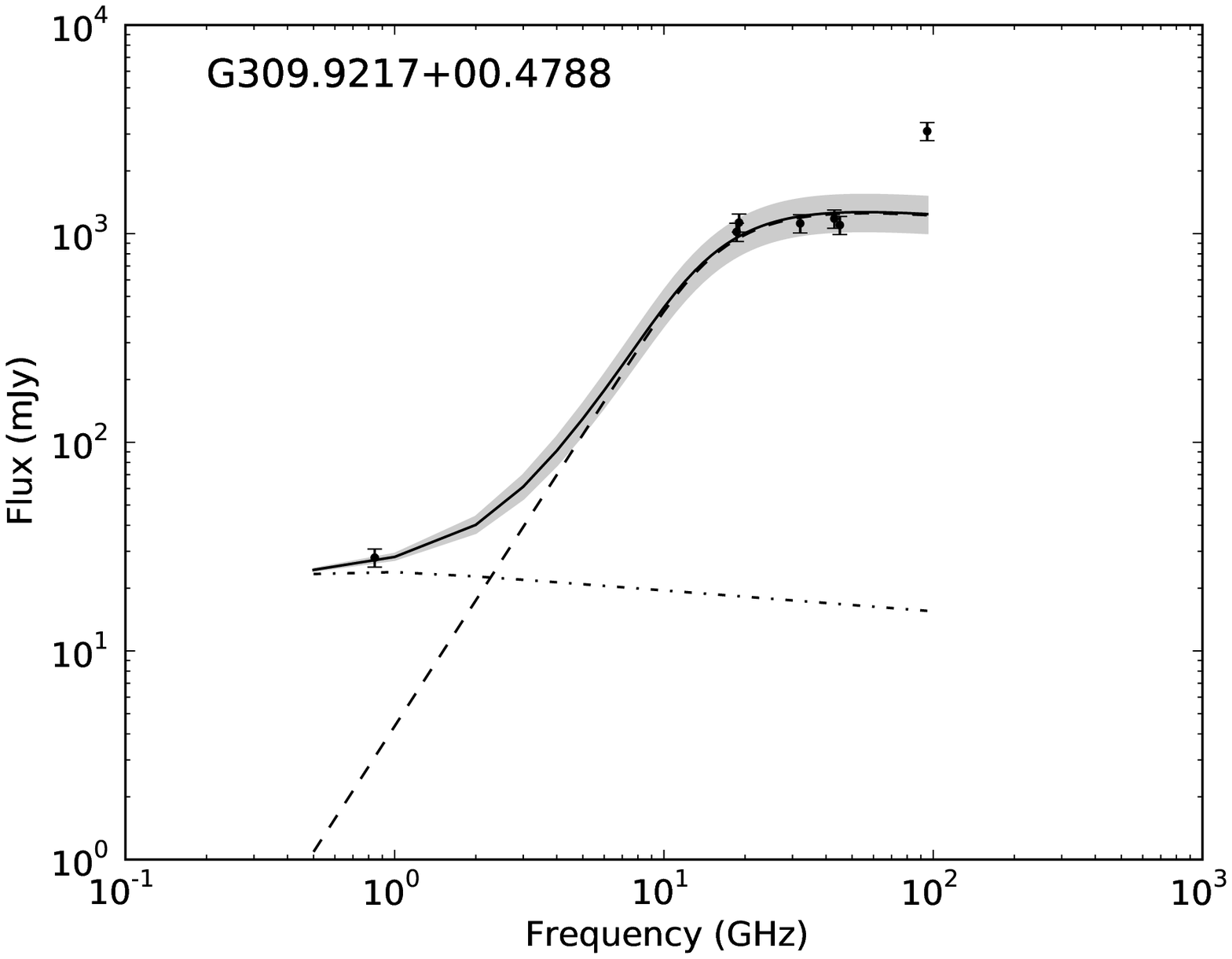}
\includegraphics[width=7cm]{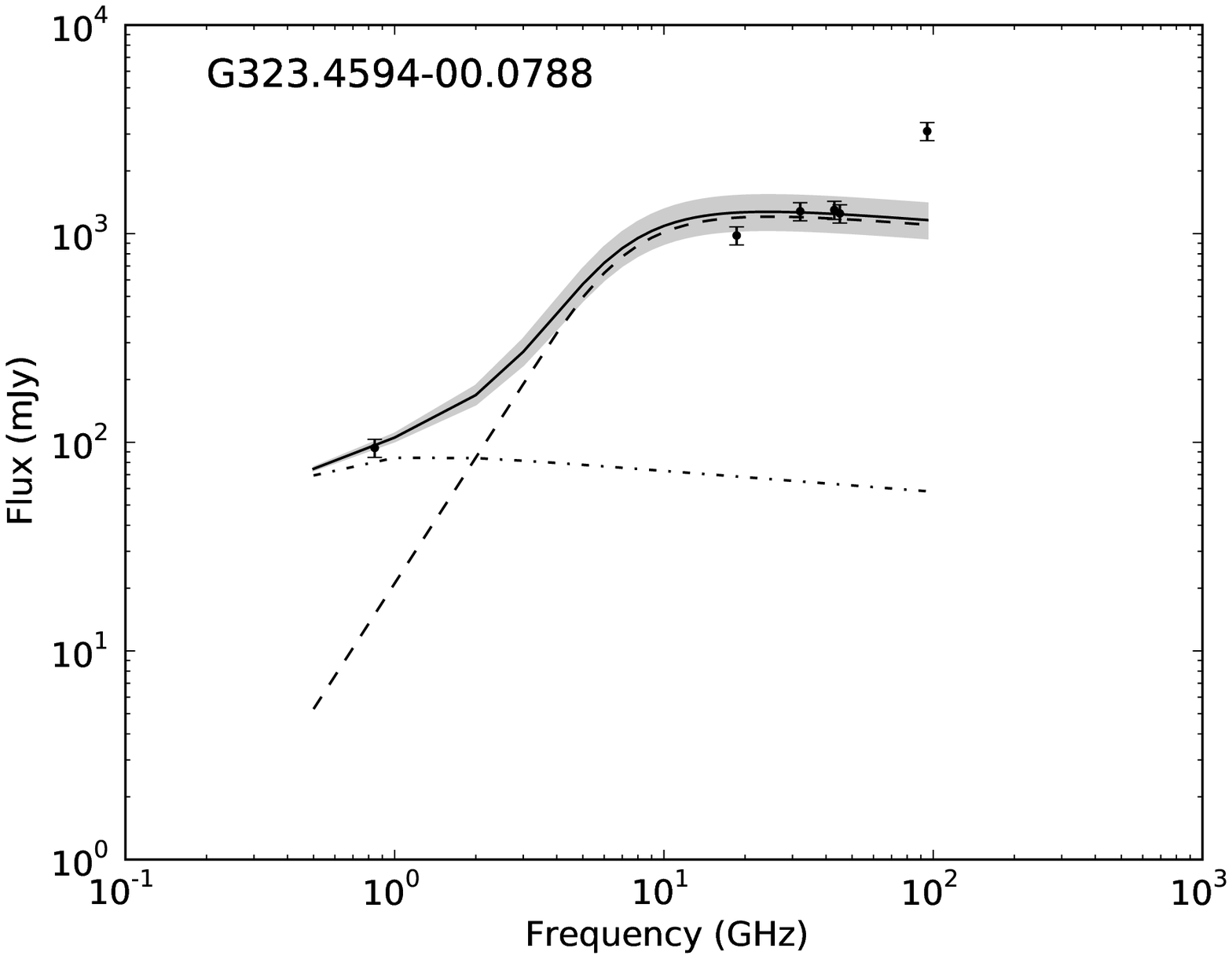}
\includegraphics[width=7cm]{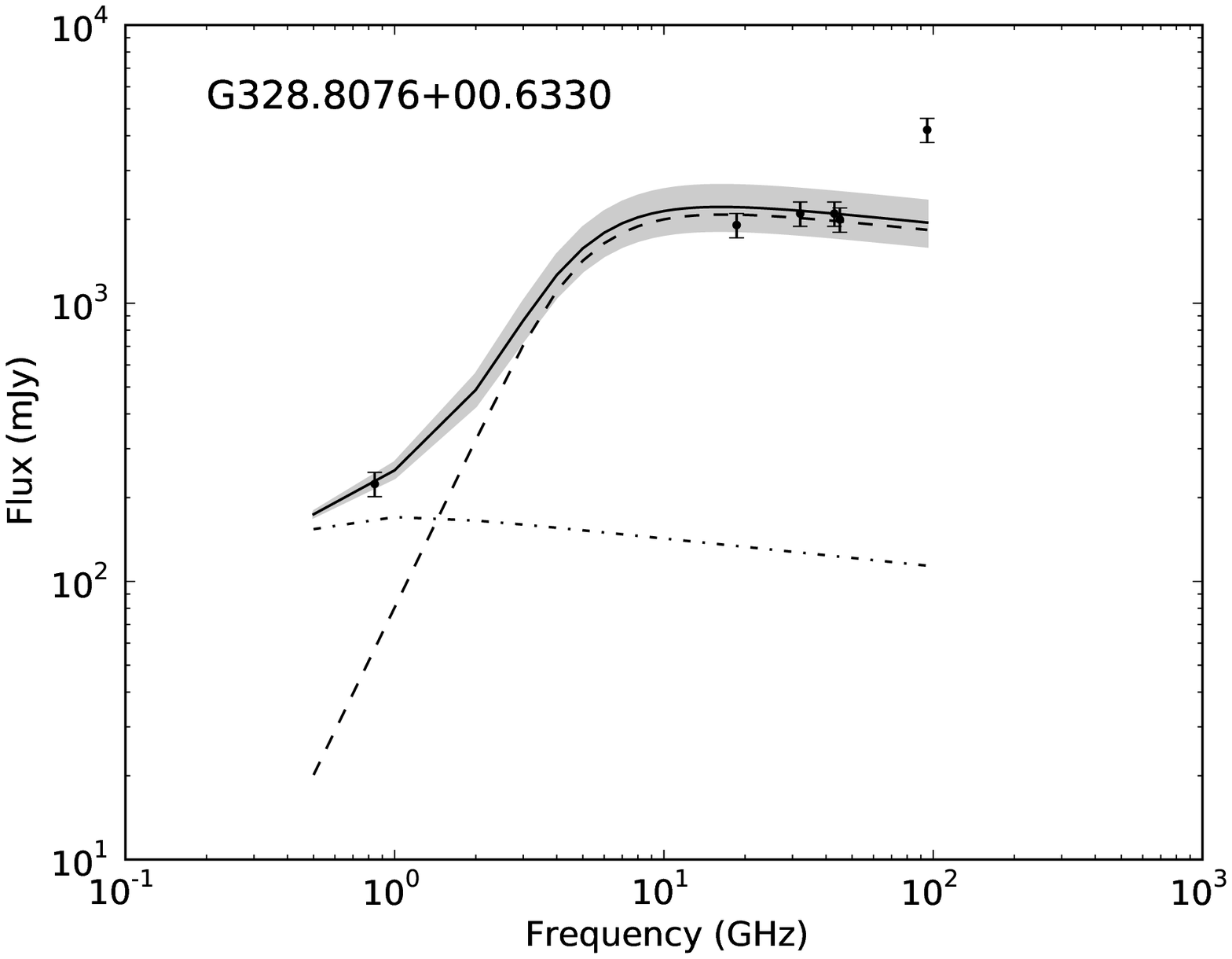}
\includegraphics[width=7cm]{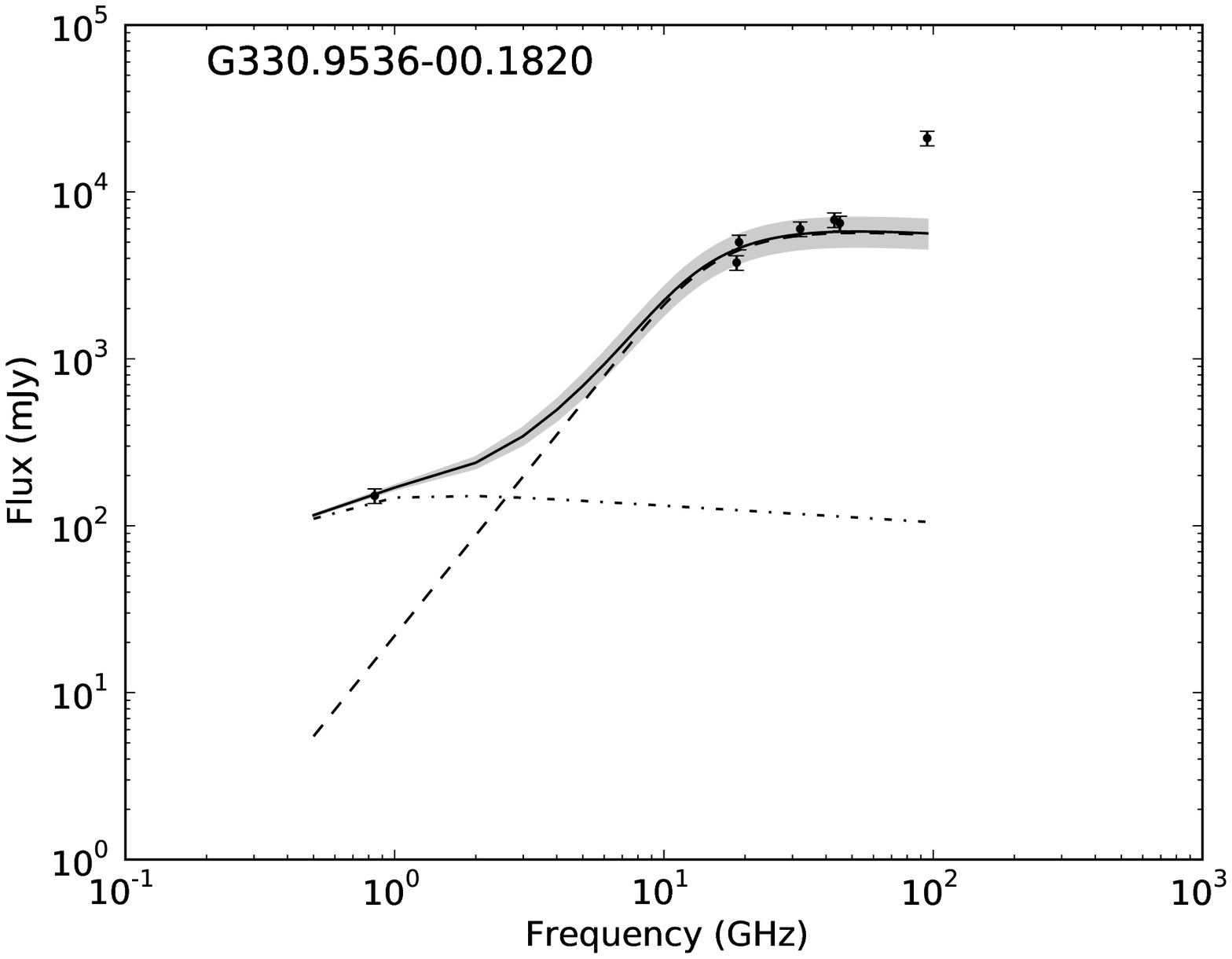}
\includegraphics[width=7cm]{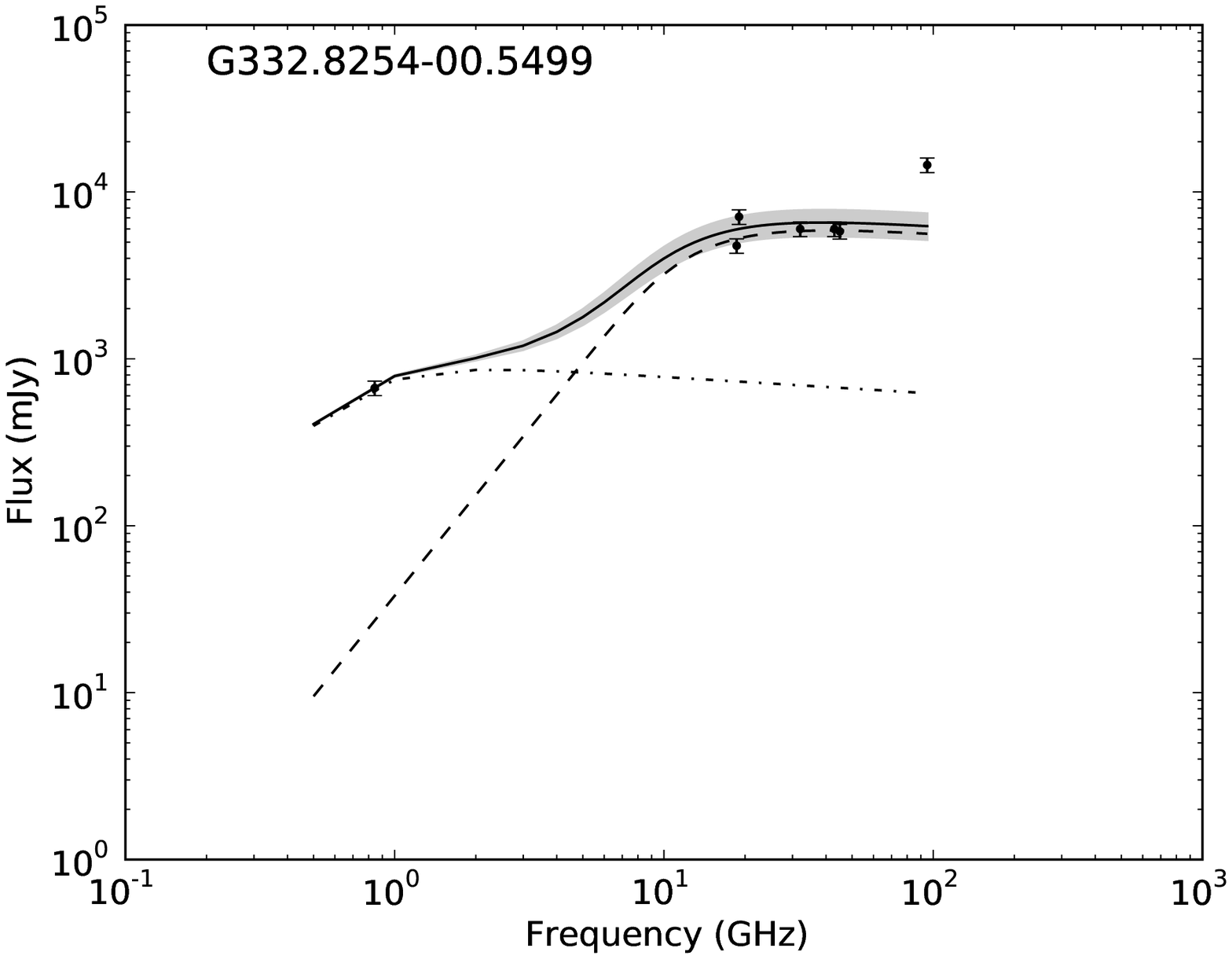}
\caption{Spectral energy distributions and best fit models for our sample. 
The dashed line shows the fit to flux density points between 1~GHz and 45~GHz.
The dotted line shows the additional component required to account for the 843~MHz
flux density. The solid line represents the sum of the two components, with grey scale 
showing the range in fits allowed by a 10\% change in measured source size.}
\label{f_seds}
\end{figure*}

The spectral energy distributions (SED) for each of our eight objects are shown 
in Fig.~\ref{f_seds}.
The flux densities increase from 843~MHz to 95~GHz (or the highest frequencies for sources 
not observed at all bands).
Table~\ref{t_models} tabulates quantities derived from constant density models fit to 
the SEDs over the range 843~MHz to 45~GHz.
For each of the eight sources we give the observed recombination line temperature and line 
width, followed by the emission measure (EM), the predicted linear size, mean electron density, 
A$_V$, and our classification of the type of HII region based on all these data 
(U: ultra-compact or H: hyper-compact).

\changes{We have modeled each source's SED to find the best estimate for its physical
parameters. The models also allow the flux density in the Planck Low Frequency Instrument 
(LFI) and High Frequency Instrument (HFI) bands to be estimated by interpolation or extrapolation.}
To model the SEDs we fit the component visibilities, measured source sizes and the 
recombination line electron temperature.
We used the standard model for the electron density of an HII region, given by 
\citet{mezger67}. 

We estimated each source's distance as described in Section~\ref{s_distance}. 
All flux densities were calculated by fitting the visibilities and extrapolating
back to zero spacing. The errors associated with this process are shown on each point in 
Fig.~\ref{f_seds}. Any contribution from large scale diffuse emission 
was excluded. The source sizes were calculated by fitting a Gaussian to the 19~GHz 
visibilities. These sizes are 
accurate to $\pm10\%$, and the range of allowed models is represented by 
grey shading in Fig.~\ref{f_seds}.
If the output from the snapshot observations of cuts in hour angle were inconsistent with 
circular symmetry then the geometric mean size was used.  

Given the angular size constraints from the visibility functions at all frequencies 
we have only one free parameter, the emission measure, which was calculated by.
assuming free-free emission and
fitting uniform density SEDs to the observed flux density distribution 
in the frequency range 1~GHz to 45~GHz. A second component (shown as a dotted line for each
model in Fig.~\ref{f_seds}) was added to demonstrate
that the higher than expected flux density measurement at 843~MHz can be explained by
a diffuse component ($\sim10\arcsec$ in size) surrounding the core of the UCHII region.
\changes{Although this more diffuse component is not imaged directly, it is well constrained 
by the observations.  It must be smaller than $\sim$45\arcsec or it would appear 
resolved in the MPGS-2 image and it must be larger than $10-20$\arcsec or it will 
violate the 20~GHz size and flux density. The corresponding emission measures are 
between $10^4$ and a few $\times10^5$ pc~cm$^{-6}$.}

These simple models suffice to match the SEDs of G301.1366$-$00.2248 (which lacks data 95~GHz), 
G302.0321$-$00.0606 and G307.5604$-$00.5875.
However, the rest of the objects all show sharp increases in flux between 45~GHz and 95~GHz.
The most extreme case, G330.9536$-$00.1820, has a 95~GHz flux 3.5 times that measured at 45~GHz.  
This phenomenon is well established \citep[e.g.][]{gibb07} though not well understood.
Attempts to fit these by purely-free-free components fail to match the observed source 
sizes by requiring a smaller angular size at 95~GHz, whereas the visibilities indicate a 
diameter similar to that of the 19~GHz flux. 

There is a tendency in the literature to characterize all HCHII region spectra as approximately
$\nu^{+1}$ but there are significant variations even in our small sample.  Nonetheless, 
the values are suggestive of the spectra found for (proto)planetary nebulae,
and ionized stellar winds in general. 
Several groups have proposed explanations, interpreting the steep spectra as 
thermal emission either from constant density cores \citep{olnon75}, or a constant 
velocity wind with an r$^{-2}$ density distribution \citep{panagia75,wright75}.

\changes{Further refinements were introduced by Marsh who proposed truncation of 
a constant velocity inverse-square stellar wind flow \citep{marsh75},
and acceleration of the stellar envelope by radiation pressure
on dust grains in the wind \citep{marsh76}. All these 
approaches require the size of the unit optical depth surface to be 
frequency-dependent. Both mechanisms would steepen the spectral energy 
distribution of the radio emission, which constituted the essential 
constraint for modeling single dish data. This aspect of early models was 
successful. 
However, when one also has multi-wavelength data including visibilities 
from synthesis imaging at a range of antenna separations, the spatial 
scale of the radio-emitting envelope has further constraints and may 
not be consistent with an analysis of single dish observations.}

\citet{sewilo04} allude to the possibility of clumping of the gas on an unresolved 
scale and this was discussed by \citet{ignace04} and \citet{gibb07}.
If the 95~GHz (3~mm) free-free emission had a very small filling factor over the 
entire measured size then this component would become optically thick below 45~GHz (7~mm) 
and one could fit the rise at 95~GHz without violating the size constraint.  
\changes{It is possible that a thin shell or even the sharp edge of a shell, could provide 
the required small scale structure but more complete high resolution observations would 
be needed to confirm this.}

Thermal emission by dust is more commonly invoked to account for the 95~GHz excess,
particularly for sources where dust emission is shown to be a natural extension of 
the associated MIR and FIR spectral energy distribution \citep[e.g.][]{sabbatini05}.
This could be cold dust ($\sim$30~K) mixed into the molecular core 
(Brooks et al. in preparation)
or warmer dust, heated by the embedded star(s) forming in the core, and expelled 
from the core into the envelope by a radiatively driven stellar wind.

\citet{lizano08} argues that heating of grains by collisions with gas to their 
sublimation temperatures could not occur unless the gas densities exceeded
$10^9{\rm cm}^{-3}$.  We have also been able to model the high frequency excesses
by a shell of hot dust ($\sim800$~K) for which we require similar high mean 
densities $2\times10^9{\rm cm}^{-3}$.  Full radiative transfer modelling would be
needed to verify this scenario.  This should be guided by FIR observations,
preferably with significantly higher resolution than IRAS to avoid confusion 
in the Galactic plane caused by unassociated point sources and widespread diffuse emission.
MIPSGAL 70-$\mu$m images will be useful when they become available. 
The modelled dust contributions cannot be excluded solely from their implied temperatures.
Our SEDs have the great merit of fitting all the data for a given source and hence
we can estimate flux densities to satisfy the Planck LFI calibration requirements
through interpolation.

The 843~MHz flux densities predicted from our simple models were lower (by up to
a factor of 10) than our observed values. 
Assuming the excess flux densities are also produced by thermal emission then it is 
easy to match the observed values by adding an optically thin component of 
a larger size (typically $\sim10\arcsec$) but with an emission measure of only 1\% of 
the 19~GHz component.  This dominates the 843~MHz emission yet does not perceptibly 
influence the high frequency fluxes.  
However, it is not possible to determine the size, temperature and EM of this component 
uniquely from our data because the resolution of the 843MHz data is too low to resolve 
this structure.
We emphasize that this aspect of our models relates solely to the ionized volume surrounding 
the UCHII regions.
These zones also support the idea \citep{kurtz99} that every UCHII 
region is associated with an extensive low ionization halo sustained by leakage of UV
photons from the compact core.  

\section{Criteria for ultra- and hyper-compact regions}\label{s_criteria}
To discriminate between ultra-compact (UC) and hyper-compact (HC) HII regions we 
have adopted standard quantitative criteria based on a survey of the literature
\citep[see, for example][]{kurtz00,kurtz02,kurtz05,wood89a,afflerbach96,sewilo04,hoare05}.
Typical ranges for these parameters (size, mean density, emission measure and recombination
line width) are shown in Table~\ref{t_qc}.
\begin{table}
\centering
\caption{Quantitative criteria for UCHII and HCHII, summarised from the literature.}
\label{t_qc}
\begin{tabular}{lrr}
\hline
Parameter       & UCHII & HCHII \\
\hline
Size             &  $<0.1$ pc & $<0.05$ pc \\
Mean density     &  $\geq$\,10$^{4}$\,cm$^{-3}$ & $\geq$\,3$\times$\,10$^{5}$\,cm$^{-3}$ \\
Emission measure &  $\geq10^{7}$pc\,cm$^{-6}$ & $\geq$\,10$^{8}$ pc\,cm$^{-6}$ \\
Recombination line width &  $\leq40{\rm km s}^{-1}$ & $>40{\rm km s}^{-1}$ \\
\hline
\end{tabular}
\end{table}

\begin{figure*}
\includegraphics[height=6.0cm]{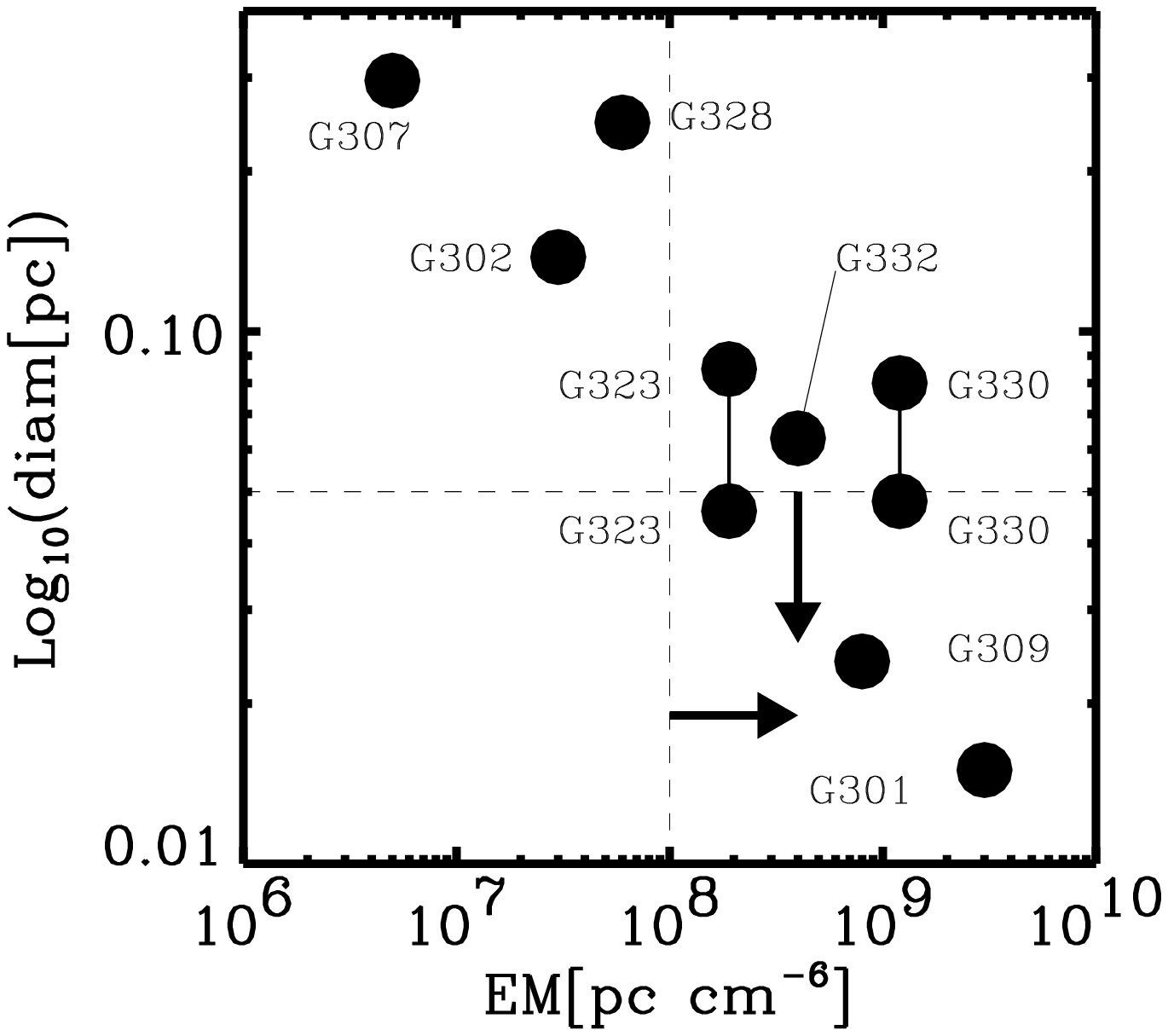}
\includegraphics[height=6.0cm]{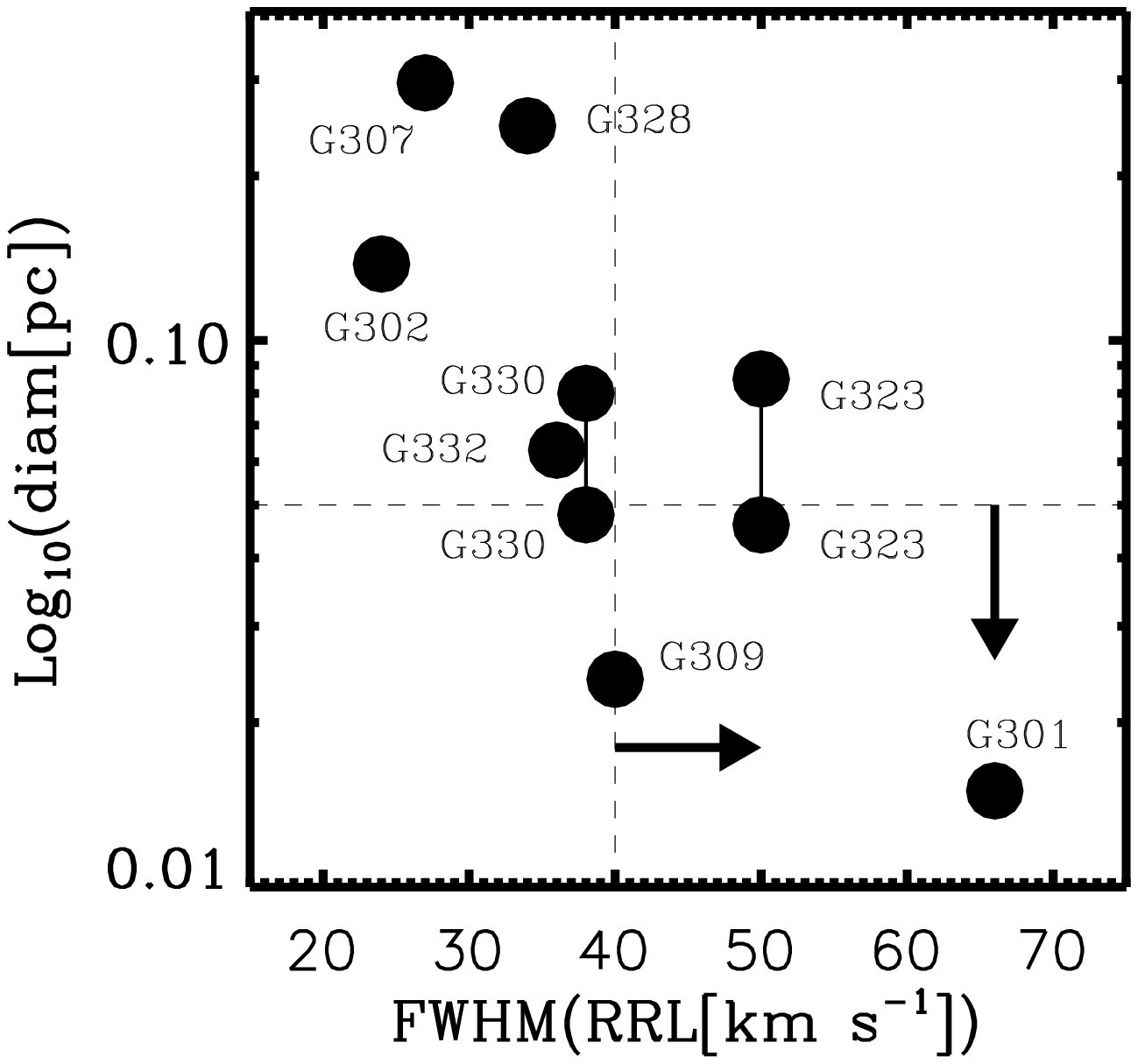}
\caption{{\it Left:} HII region diameter {\it vs.} emission measure for the eight sources in our sample. Sources are identified with a truncated version of their full name.
The two objects with unresolved distance ambiguities (G323.4594$-$00.0788 and 
G330.9536$-$00.1820) are plotted twice, corresponding to parameters derived for each 
distance, and their points are connected in 
the plot. {\it Right:} HII region diameter {\it vs.} the radio recombination line width.
The dashed lines in each figure represent the demarcation between UCHII and HCHII candidates --- 
the quadrant containing HCHII is indicated by arrows.}
\label{f_crit}
\end{figure*}

Fig.~\ref{f_crit} shows the diameter of each HII region plotted against its emission measure
(left) and recombination line width (right).
The dashed lines represent the boundaries between UC and HC regions in each attribute, 
with UCHII expected to fall towards the top left quadrant, and HCHII towards the bottom 
right quadrant.
The two sources (G323.4594$-$00.0788 and G330.9536$-$00.1820) for which we were unable to resolve the distance 
ambiguity are plotted twice (one for each distance) and connected with a line. 

If the dust distribution surrounding these nascent high mass stars were always a
disk then the orientation of that disk would govern the amount of extinction along
the line of sight rather than the mean density or EM.  G307.5604$-$00.5875 combines the 
highest measured extinction with the lowest EM, perhaps indicating that the dust envelopes 
are indeed flattened instead of spherical. No IRAS LRS spectrum is available for 
two of the three objects with the largest EMs.  This might indicate that the 
extinction to these two regions is so large that the MIR emission from them simply 
cannot escape in our direction.

It is often argued that each HC region contains only a single star, where UC regions
may have several associated stars.  Of particular interest is the complexity clearly 
shown by the Spitzer MIR images in Fig.~\ref{f_irim}. 
In addition to the associated filaments and diffuse emission, the regions we have 
observed consist of small clusters of stars, probably at different stages of early 
evolution based on their IRAC false colours. 
\changes{Even if one tried rigorously to trace every such component star in an HII 
region from one wavelength region to the next it would be a gross 
oversimplification to define an infrared SED based on the sum of 
all flux within a given complex. The underlying issue is the variation of 
beam size with IR wavelength. Constructing a NIR-to-FIR SED from 2MASS, IRAC, 
MIPS, and IRAS would involve the mingling of data from instrumental beams 
whose FWHMs change across the resulting SED from 2.3, 1.9, 6, to about 25--100 
arcsec, respectively.  No individual entity in these regions is likely to match
the shape of the total SED because the content of these beams would only 
very rarely consist of the same single entity at all wavelengths.}

\changes{Inspection of the 95~GHz (3~mm) interferometer visibilities indicate that
several of our HII regions have small structures present which are not
present in the lower frequency visibilities (or images) where they are 
presumably optically thick.
The visibilities of G323.4594$-$00.0788 clearly shows two components, with approximately
half the flux ($\pm20\%$) emitted by each component.}
Our 45~GHz data  yields only the  total flux and at 95~GHz 
we are dominated by the smaller component which we estimate as only 0\farcs7 in size.
It is likely to be another hyper-compact source.  In summary, our sample contains 2
hyper-compact sources (G301.1366$-$00.2248, G309.9217$+$00.4788), 4 ultra-compact sources 
(G302.0321$-$00.0606, G307.5604$-$00.5875, G328.8076$+$00.6330, G332.8254$-$00.5499), 
and two sources with attributes intermediate between hyper-compact and ultra-compact,
with a final classification partly dependent on resolving their distance ambiguities 
(G323.4594$-$00.0788, G330.9536$-$00.1820).

\section{Infrared Data}\label{s_mir}
\begin{figure}
\centering
\includegraphics[height=4.0cm]{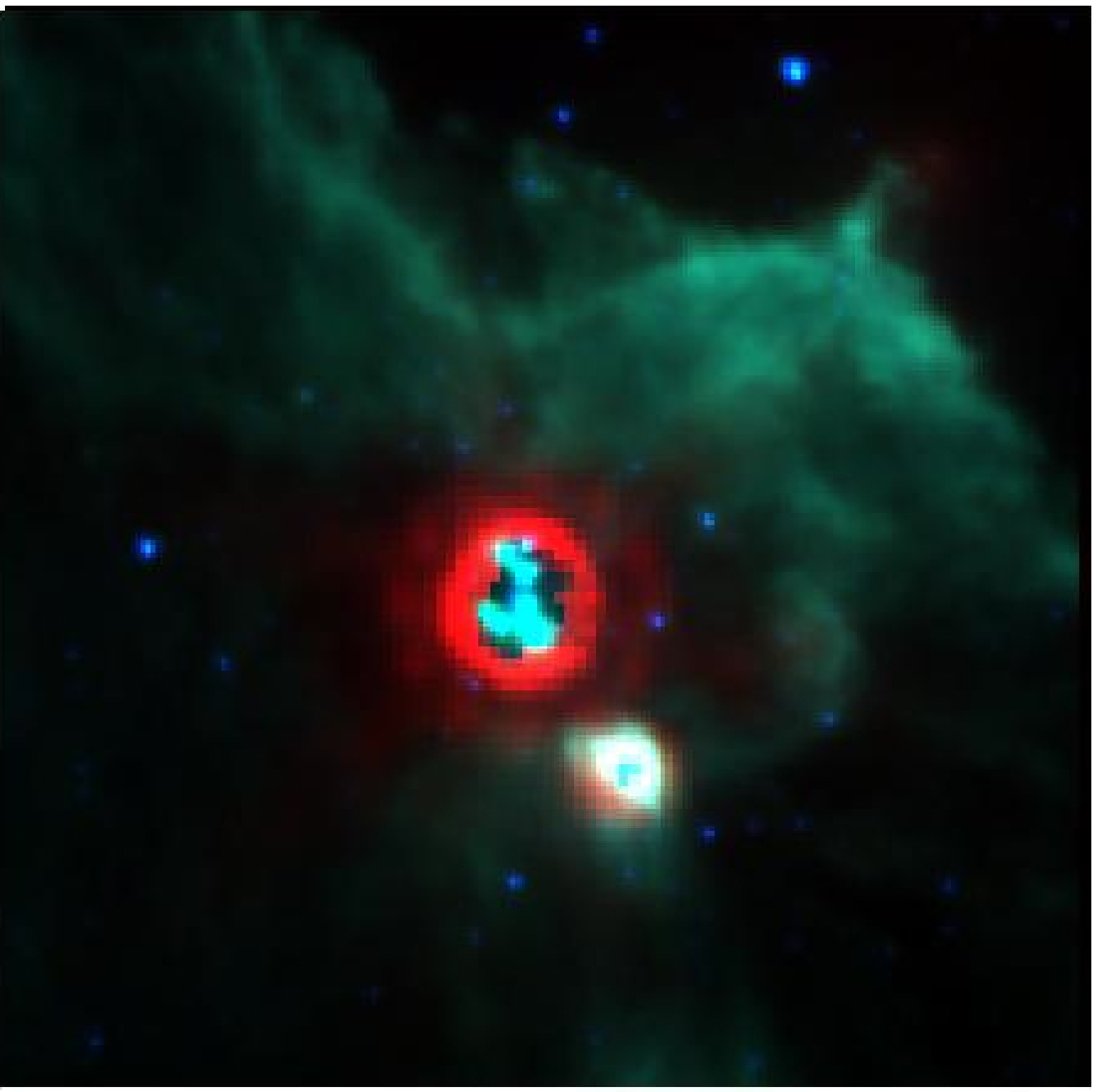}
\hfill
\includegraphics[height=4.0cm]{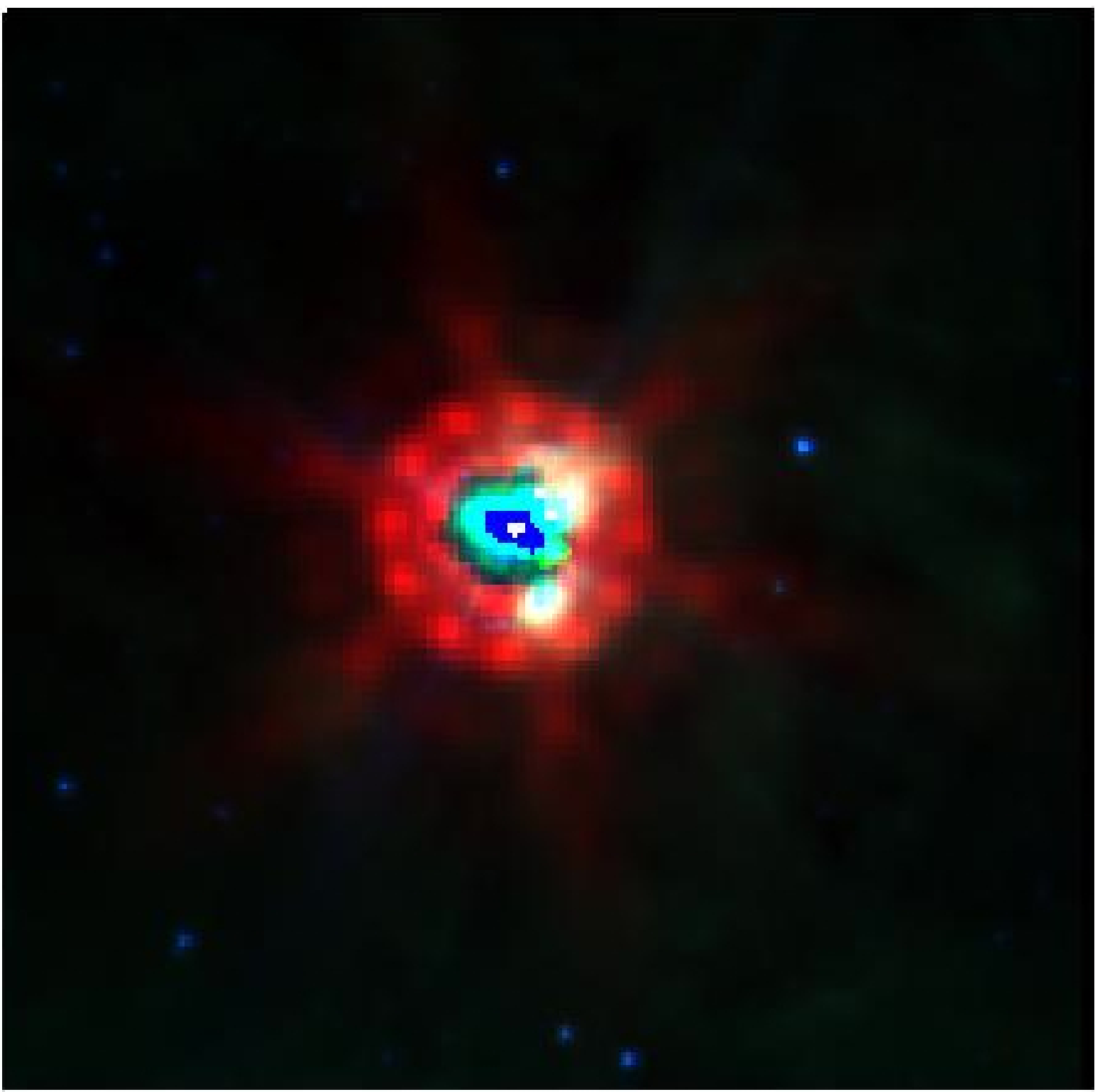}
\caption{Spitzer IRAC-MIPS false colour $24\mu$m images of G301.1366$-$00.2248 and G309.9217$+$00.4788.}
\label{f_mips}
\end{figure}
To further explore the nature of each radio-selected source we extracted images from
four Spitzer Space Telescope Legacy projects. 
Firstly, GLIMPSE-I \citep{bobben03}, which mapped the inner Galactic plane between 
longitudes $\pm10$\degr and $\pm65$\degr and latitudes $\pm1$\degr using the InfraRed 
Array Camera (IRAC) \citep{fazio04} at 3.6, 4.5, 5.8, and $8.0\mu$m.
Secondly, the later GLIMPSE-II described by \citet{churchwell07} which completed the 
coverage of the Galactic Centre region and extended the latitude range to $\pm2$\degr 
at the Centre.  
Finally, the MIPSGAL-I and MIPSGAL-II projects \citep{carey05,carey07} which matched
the GLIMPSE projects using the Multiband Imaging Photometer for Spitzer (MIPS) at 
24, 70 and $160\mu$m.

The IRAC images provide spatial resolutions between 1\farcs5 and 1\farcs9 FWHM, 
comparable with that achieved by the ATCA at 20~GHz using the 6~km baseline.  
False-colour IRAC imagery is valuable in diagnosing the nature of the IR emission 
process in a wide variety of objects.
\citet{cohen07} have discussed the values for coding IRAC images of HII regions so
that blue, green and red represent 4.5, 5.8 and 8.0\,$\mu$m, respectively.
Structures that appear yellow in the resulting 3-colour images are indicative of 
sources in which polycyclic aromatic hydrocarbons (PAHs) dominate the emission.  
By contrast, regions that appear white trace broadband thermal emission by heated 
dust, indicative of the MIR-emitting surface of the cocoon.  White regions represent 
less evolved structures than yellow regions. 

The spatial resolution of MIPS at $24\mu$m is $\sim6\arcsec$, substantially
poorer than that of IRAC. However, its ability to trace thermal emission from cool
circumstellar dust (temperature around 120~K) surrounding deeply embedded massive stars
is an important tool for isolating the youngest object in a group of sources.
We made 3-colour Spitzer images that combine IRAC 5.8, 8.0 and MIPS $24\mu$m
images as blue, green and red, respectively to define the locations of cooler
thermally emitting dust. The $24\mu$m images were from the 2007 October release of 
enhanced MIPSGAL images, described by \citet{carey09}. The $24\mu$m peaks always coincide
with the white IRAC sources and PAH emission would appear turquoise in the IRAC-MIPS
combined colours as demonstrated in Fig.~\ref{f_mips}.

\subsection{Comparison of high resolution MIR and radio observations}
\begin{figure*}
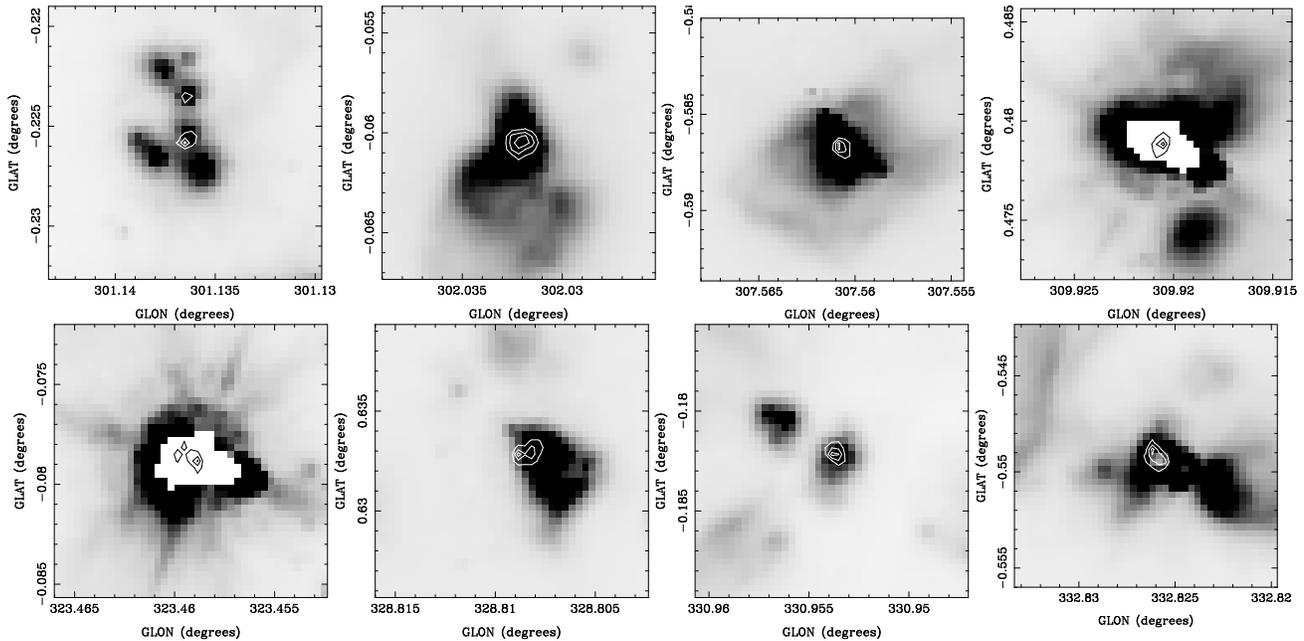

\centering
\includegraphics[width=4.2cm,angle=270]{images/f30114m022_4.uv.ps}
\includegraphics[width=4.2cm,angle=270]{images/f30203m059_4.uv.ps}
\includegraphics[width=4.2cm,angle=270]{images/f30756m059_4.uv.ps}
\includegraphics[width=4.2cm,angle=270]{images/f30992p048_4.uv.ps}
\includegraphics[width=4.2cm,angle=270]{images/f32346m008_4.uv.ps}
\includegraphics[width=4.2cm,angle=270]{images/f32881m063_4.uv.ps}
\includegraphics[width=4.2cm,angle=270]{images/f33096m018_4.uv.ps}
\includegraphics[height=4.2cm,angle=270]{images/f33283m055_4.uv.ps}
\caption{Overlays of MIR and high resolution 19~GHz images demonstrating the
relationship between radio emission and the MIR structure of these regions.
The white contours (or black contours in the cases where the MIR image
is saturated) of radio emission show 3 levels, equally spaced 
between $5\sigma$ and the image peak. Sources shown from left to right then top to bottom are:
G301.1366$-$00.2248; G302.0321$-$00.0606; G307.5604$-$00.5875; G309.9217$+$00.4788; 
G323.4594$-$00.0788; G328.8076$+$00.6330; G330.9536$-$00.1820; and G332.8254$-$00.5499.}
\label{f_radmir}
\end{figure*}

Our high resolution 19~GHz images have almost the same spatial resolution as the IRAC
images and hence are invaluable for associating the dominant radio emission of a region
with specific MIR components.  
Fig.~\ref{f_radmir} shows our 19~GHz continuum images (as contours), 
overlaid on grey-scale $8\mu$m images, which enables us to locate the origin of the 
radio emission.  These indicate that the radio peaks match the white IRAC sources and 
the $24\mu$m peaks.

For example, the most well studied source in the set, G301.1366$-$00.2248 
\citep[see][]{henning00} appears to consist of about 6 separate MIR components, some diffuse.
Yet radio emission is found from only two of these objects, and the stronger radio component
is far brighter than the second radio source in this complex.  This is important
because this region also has the highest recombination line width in our sample.  This
66\,km\,s$^{-1}$ line width is intrinsic.  It could not arise simply by overlaying lines
from the two radio components.

\subsection{MIR morphologies}
\begin{figure*}
\centering
\includegraphics[width=4.3cm]{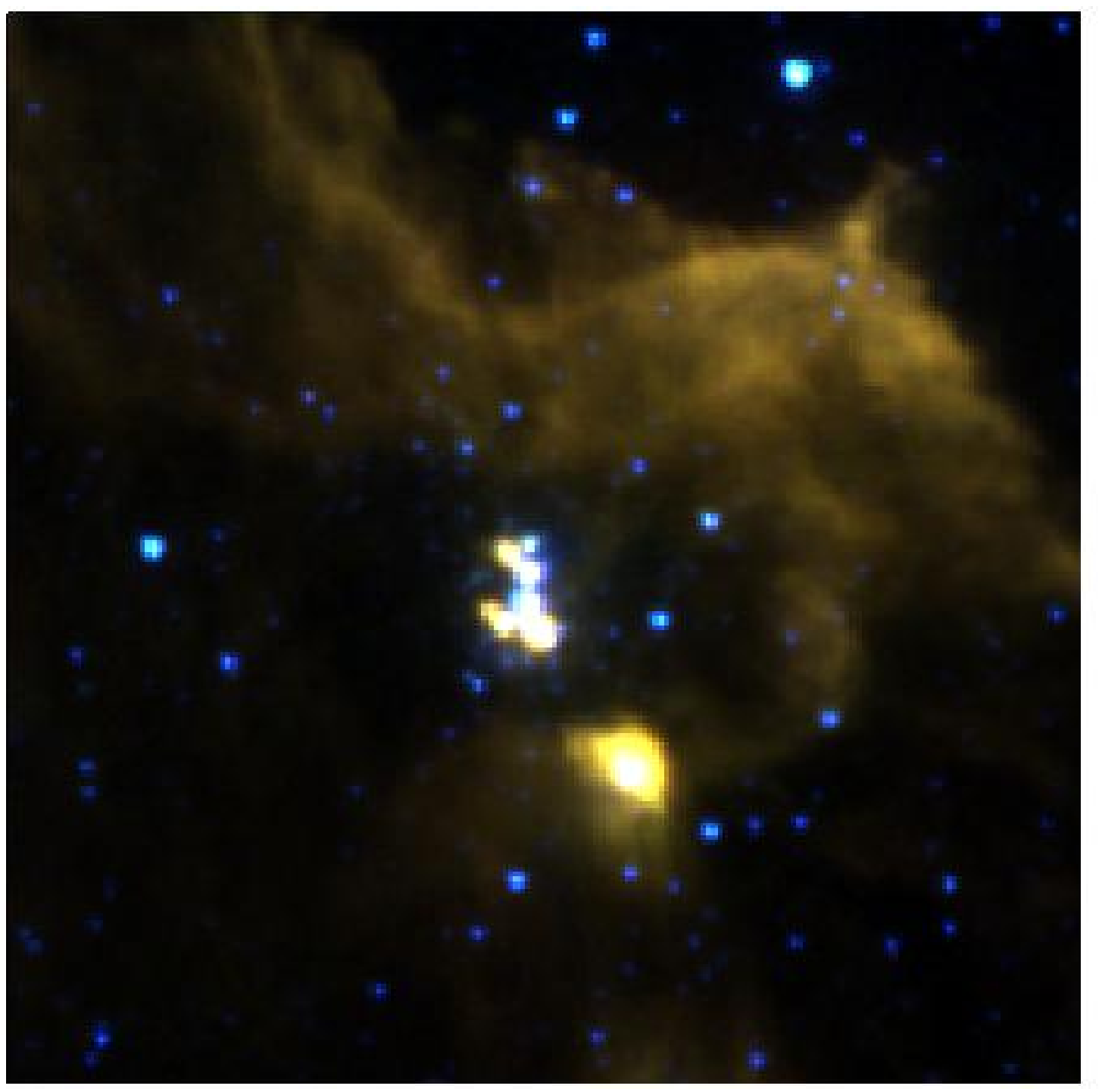}
\includegraphics[width=4.3cm]{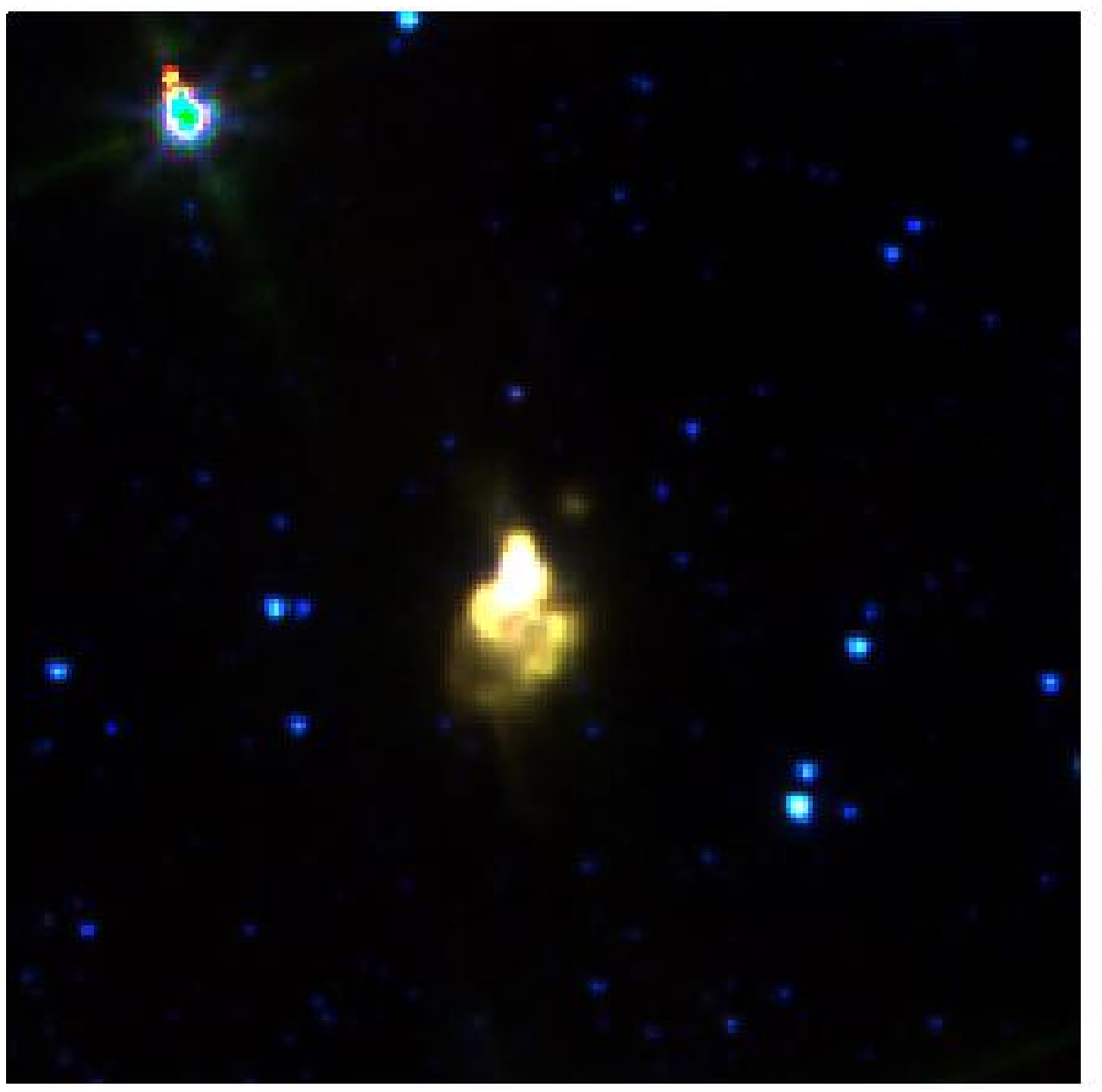}
\includegraphics[width=4.3cm]{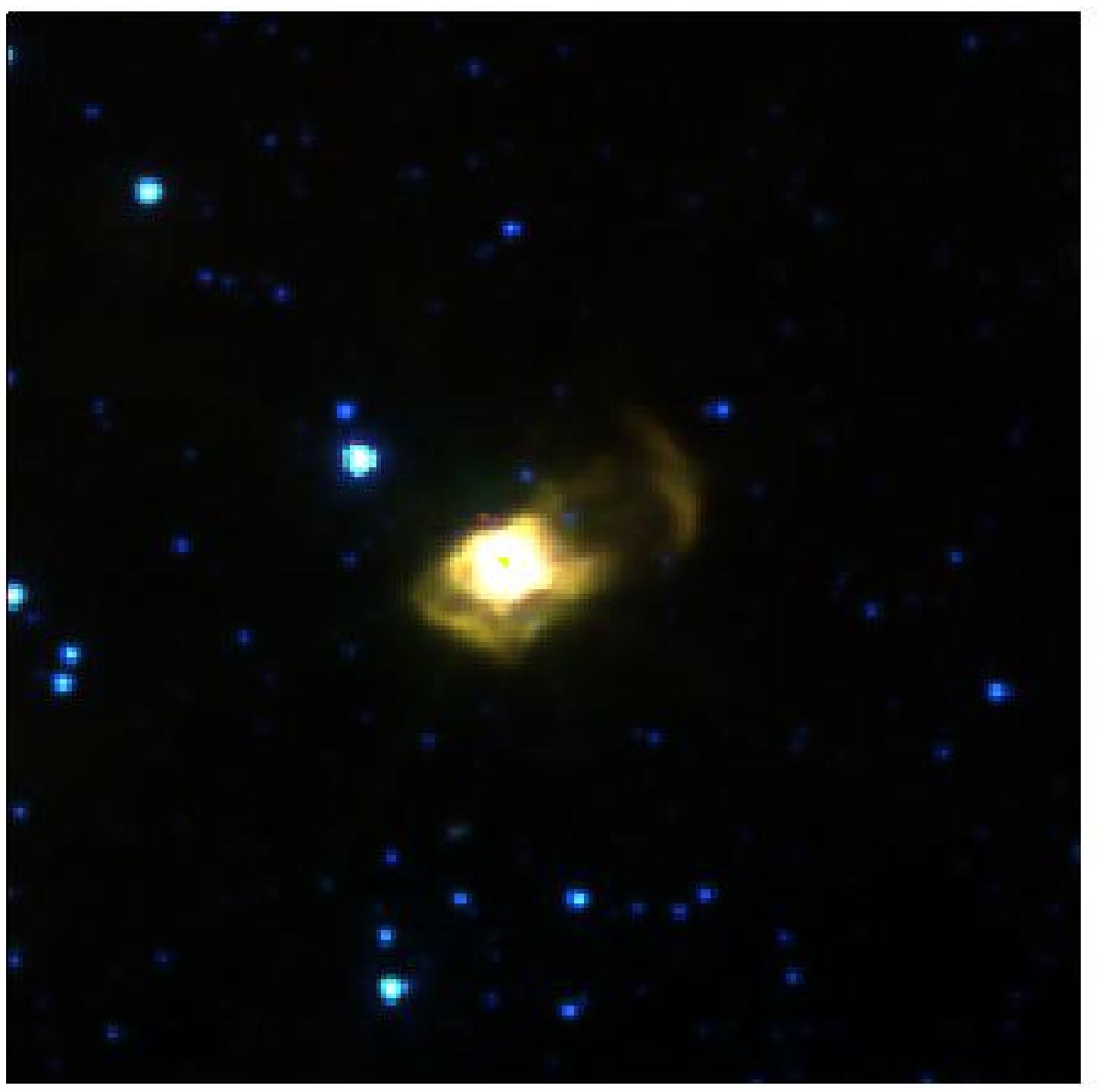}
\includegraphics[width=4.3cm]{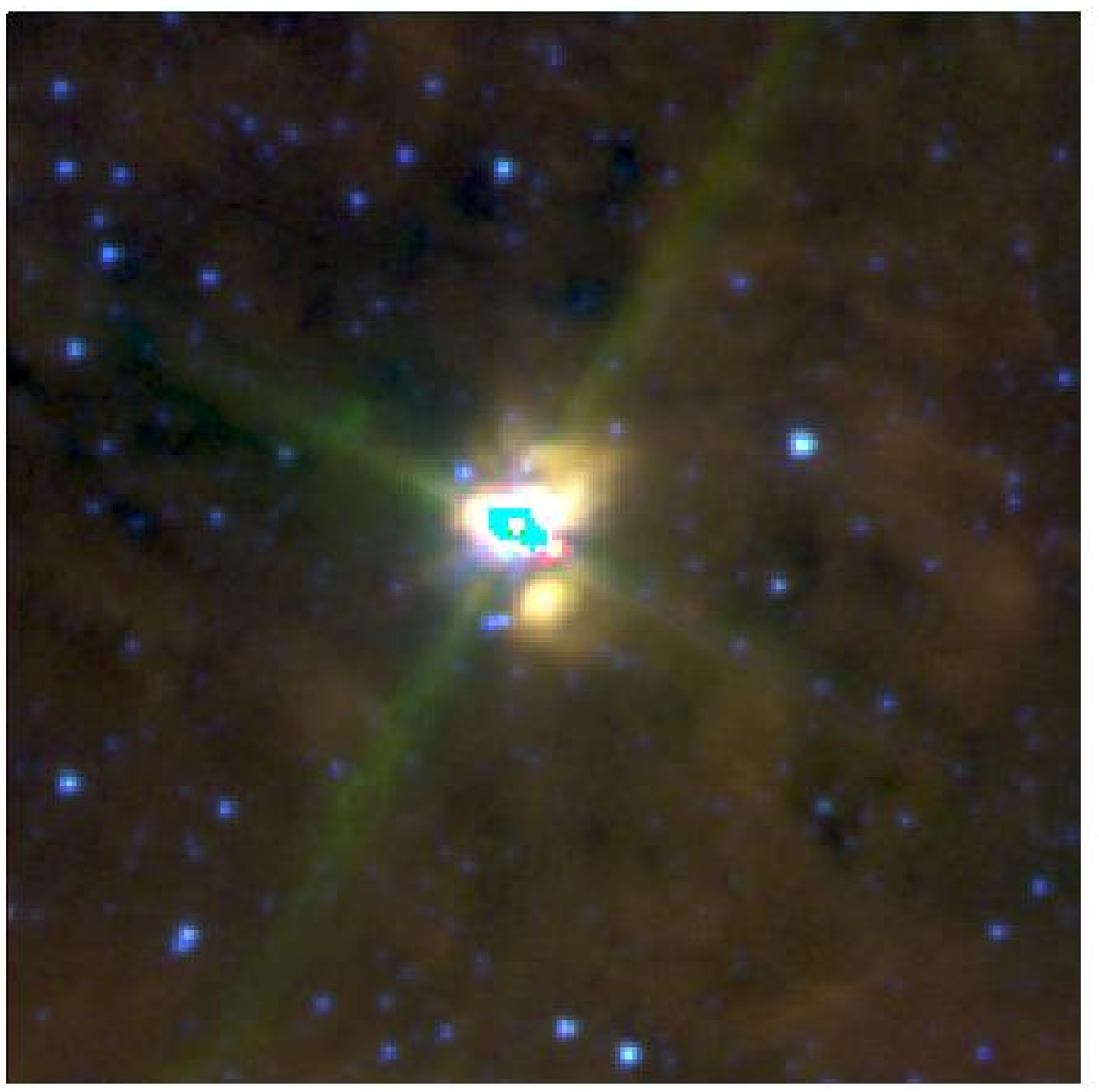}
\includegraphics[width=4.3cm]{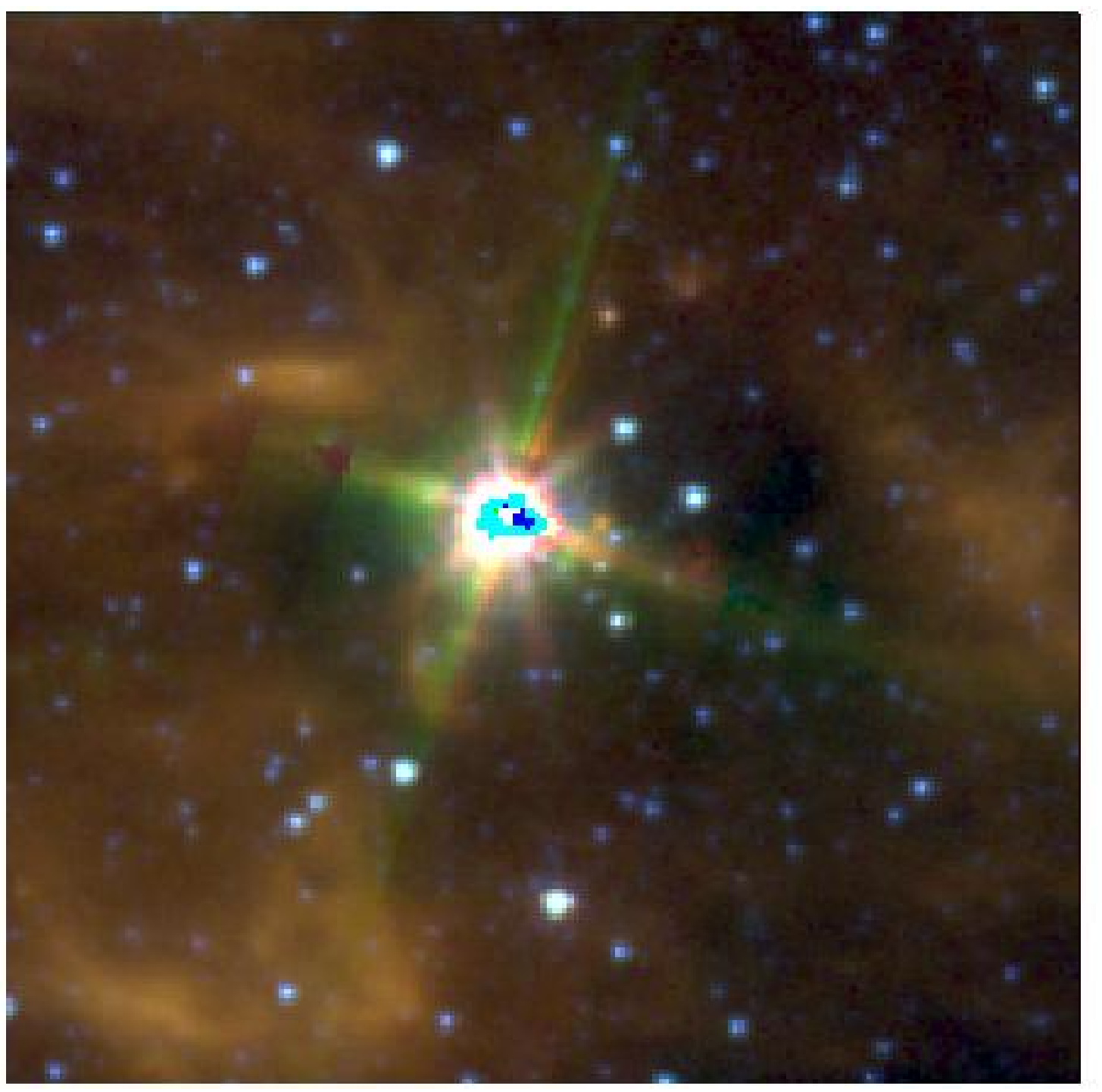}
\includegraphics[width=4.3cm]{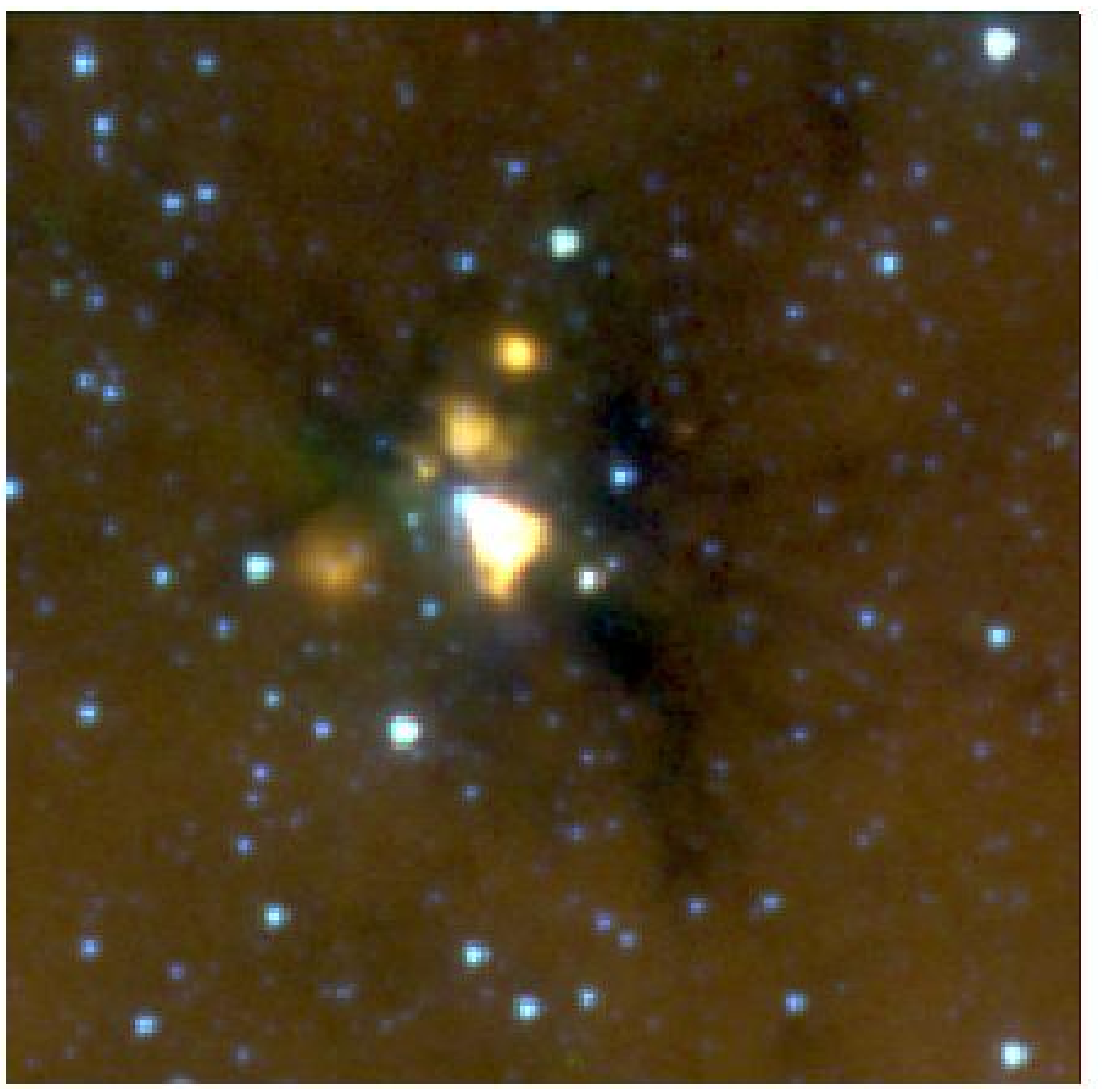}
\includegraphics[width=4.3cm]{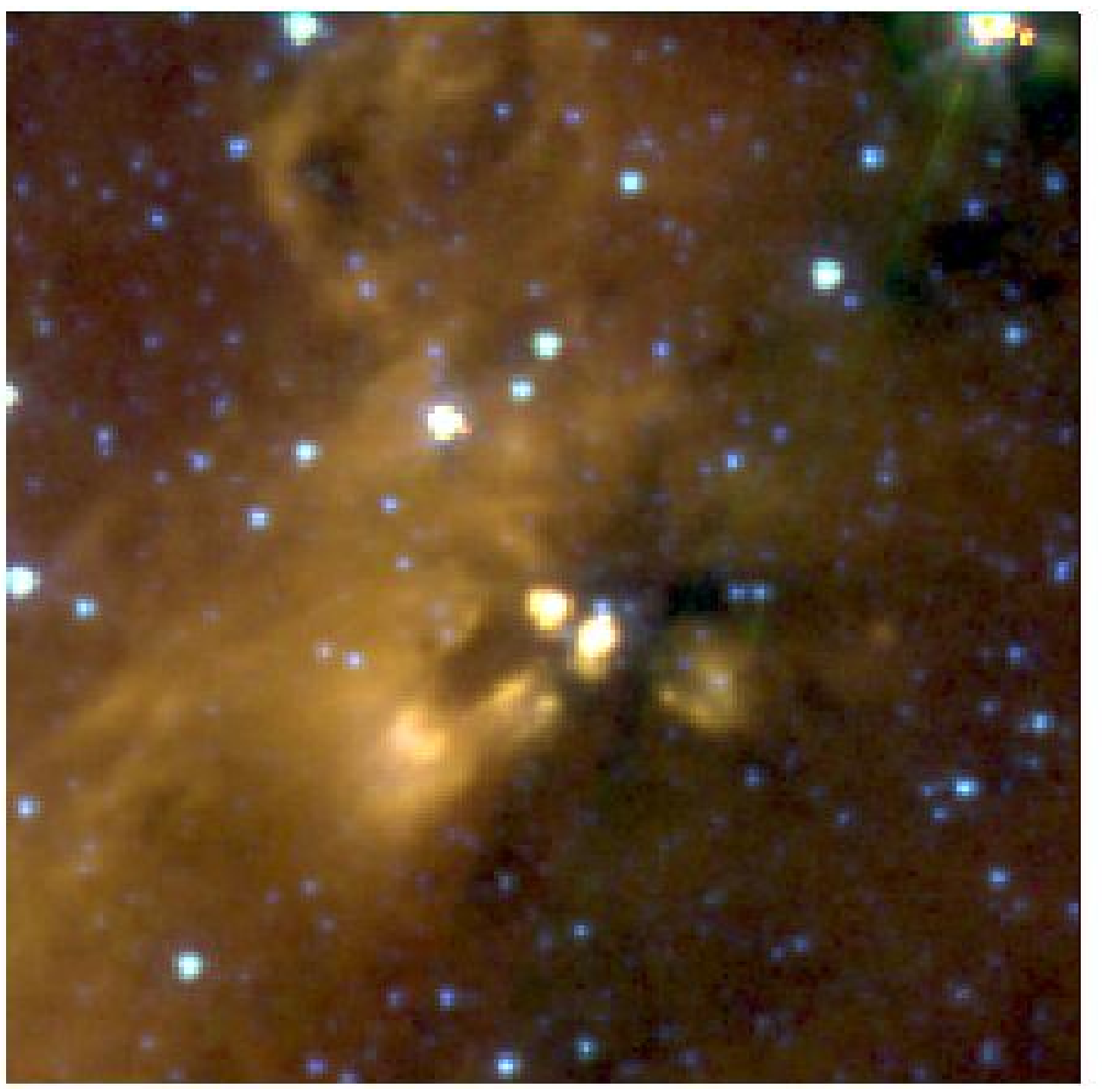}
\includegraphics[width=4.3cm]{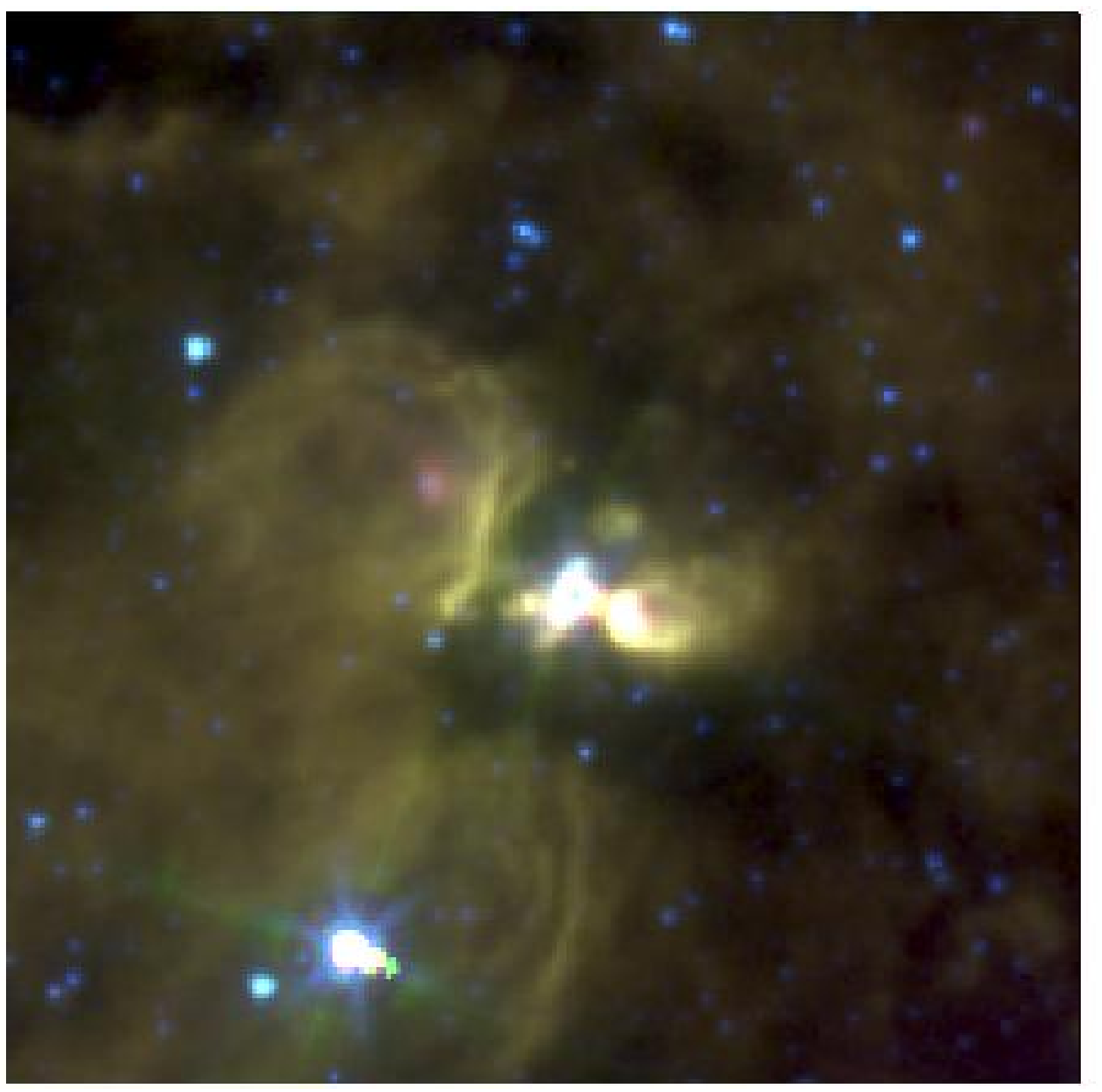}
\caption{MIR false-colour images of UCHII regions.
For the Spitzer images we used blue, green, red for images at 4.5, 5.8, 8.0\,$\mu$m bands,
respectively. Sources shown from left to right then top to bottom are:
G301.1366$-$00.2248; G302.0321$-$00.0606; G307.5604$-$00.5875; G309.9217$+$00.4788; 
G323.4594$-$00.0788; G328.8076$+$00.6330; G330.9536$-$00.1820; and G332.8254$-$00.5499.}
\label{f_irim}
\end{figure*}

High spatial resolution MIR views of our sample are provided by the false-colour
IRAC images shown in Fig.~\ref{f_irim}.
From these, one can appreciate that: (i) MIR and radio
morphologies of these sources are totally different; (ii) our sample shows a 
diversity of MIR structures; (iii) most objects consist of clusters of MIR point 
sources or small extended sources, often in association with diffuse emission and/or 
separate bright MIR sources.   Most of the point
sources that constitute these small groups are yellow (indicative of PAH emission) but
a small number (typically only one) of the cluster members are white, suggesting thermal
radiation by warm dust grains which emit in all three of the IRAC bands. 
These white objects are always the brightest elements of the MIR groups. When such an
ultra-compact MIR object is detected it is always
closely accompanied either by other, fainter, sources or diffuse emission, or both.
Turquoise cores arise because of saturation in the IRAC arrays. Red, orange or yellow
diffuse emission is often observed around the small groups, again suggesting widespread
PAH emission as \citet{cohen07} found for spatially extended HII regions.

The MIR counterparts of G309.9217$+$00.4788 and G323.4594$-$00.0788 contain obvious
indications of an unresolved source as evidenced by the diffraction spikes from the
Spitzer telescope as seen in the IRAC images.
There is no relationship between MIR and radio morphologies for HII regions in
general \citep{cohen07a} and there appears to be none for HCHII and UCHII regions either.

\subsection{Extinction within UCHII regions}\label{s_extinct}
\begin{figure}
\centering
\includegraphics[width=7.0cm]{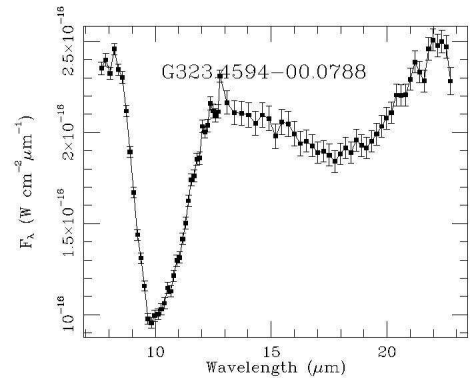}
\caption{IRAS LRS spectrum of G323.4594$-$00.0788 showing the deep silicate absorptions.}
\label{f_lrs}
\end{figure}

We have examined false-colour images based on 2MASS $JHK$ bands to aid interpretation
of the MIR images. However, very few near-infrared (NIR) counterparts to the UCHII 
regions are found, presumably due to heavy obscuration. In order to test this presumption
and to estimate the total line-of-sight extinction toward our sources we have sought
spectra from the complete IRAS LRS database
of 171\,000 spectra from 7.7$-$22.7\,$\mu$m.
All spectra were recalibrated by \citet{cohen92}.  Fig.~\ref{f_lrs} shows a representative 
spectrum for the source G323.4594$-$00.0788. The most obvious features are deep 
amorphous silicate absorption bands near 10 and $18\mu$m.  These arise as light from 
the embedded MIR-bright  central star suffers heavy extinction along the line-of-sight 
within the dusty circumstellar cocoon.

The $10\mu$m optical depths of the silicate features can be converted into measures
of the associated extinction by multiplying by $18.5\pm1.5$ \citep{RA84}.
In several regions the silicate absorption at $10\mu$m is so severe that the LRS
data fall to zero.  In such spectra the $18\mu$m absorption band is still readily detected 
and we estimate a median of 5.3 for the ratio of $\tau_{10\,\mu m}/\tau_{18\,\mu m}$ for a much
larger sample of HII regions than are represented here. In this way we have
derived line-of-sight extinctions for 7 of the 8 sources in this paper.  The 
expectation is that almost all of this extinction arises  solely within the dusty 
shell that still envelops each of these regions, rather than in the intervening interstellar
medium.  

\section{Conclusions}\label{conc}
We present a sample of eight hyper- and ultra-compact HII regions, identified as part of
a blind survey for ultra-compact HII regions. These are part of a larger set of 46 objects,
identified on the basis of their 20~GHz to 843~MHz radio flux density ratio. These sources are 
some of the brightest compact sources in the Southern sky at 20~GHz. The basis for
this work are two new surveys of the southern sky --- the Australia Telescope 20~GHz survey
and the second epoch Molonglo Galactic Plane Survey at 843~MHz.

We have classified each source as hyper- or ultra-compact on the basis of their emission
measures, sizes, and radio recombination line width, resulting in 2 HCHII, 4 UCHII and 
2 sources that are undetermined between these classifications because of an ambiguity 
in their kinematic distances. 
\changes{Using MIR and radio images at comparable resolution we have shown that,
on the smallest scales that we can explore, the MIR morphologies of 
these sources are completely different from their radio morphologies.}
The MIR images show that UCHII regions are rarely
simple, and what appears to be a single object at lower resolution may actually be a complex
cluster of young stars. The high frequency radio emission is useful for picking out the 
youngest objects in these clusters.

We have modelled the spectral energy distributions with a uniform density model assuming
free-free emission. A majority of the objects show excess emission at the highest 
frequency, with the flux density at 95~GHz up to three times the flux density at 45~GHz.
The confirms similar observations of young stellar objects by \citet{gibb07}. To explain 
this excess we require a more sophisticated model than the one used here. For example,
clumpiness of the gas on unresolved scales could produce the observed flux densities without
violating the size constraints. Another alternative is the presence of cold dust mixed
into the molecular core, or warmer dust formed in the core and then expelled into the 
envelope by a stellar wind. To investigate these possibilities higher resolution imaging
in radio and the FIR are required.

\subsection{Suitable calibration sources}
An important goal which we have attempted to address was to identify calibrators so
that Plank's LFI could monitor the Galactic plane foreground emission.  
Our observations span the range from 0.8 to 95\,GHz, covering the entire range of wavelengths 
for LFI (30, 44, 70\,GHz).  To extend to the 100\,GHz channel of the HFI requires only minimal
extrapolation of our models of these HII regions. Our objects were chosen on the basis of 
being compact and isolated (based on MGPS-2 observations) and as such are candidates for
Planck calibrators. 

In addition, these sources may be useful as 45~GHz (7~mm) and 95~GHz (3~mm) primary flux calibrators
or secondary calibrators) for the Australia Telescope Compact Array. A subset of these sources 
are now being monitored as part of the C007 and C2050 calibration programs.
It should be noted that the sources with excess emission at
95~GHz can not be explained by current models, and hence high frequency flux densities can
not be extrapolated from our SEDs.

\section*{Acknowledgments}
We thank the anonymous referee for providing suggestions that helped improve
the clarity of this paper.

We are grateful to Dr. David Frew for providing information on the greatest distance
that one might detect H$\alpha$ emission from these sources, and to Dr. Katherine Newton-McGee
for reducing the C007 observations. We are also grateful to Dr. Jamie Stevens for carrying
our observations as part of the C2050 calibration program at the ATCA.

TM acknowledges the support of an ARC Australian Postdoctoral 
Fellowship (DP0665973).
Partial support for MC's participation in this work under the {\it Spitzer} 
Space Telescope Legacy Science Program, was provided by NASA through contract
1259516 between UC Berkeley and the Jet Propulsion Laboratory, California Institute of
Technology under NASA contract 1407. MC is also grateful for support from the School
of Physics in the University of Sydney through the Denison Visitor program, and from
the Distinguished Visitor program at the Australia Telescope National Facility
in Marsfield. RDE is the recipient of an Australian Research Council
Federation Fellowship (FF0345330) which also provided travel support for MC.

The MOST  is owned and operated by the University of Sydney, with support
from the Australian Research Council and Science Foundation within the School of
Physics. 

This research made use of data products from the Midcourse Space
eXperiment.  Processing of the data was funded by the Ballistic
Missile Defense Organization with additional support from NASA's Office of
Space Science. This research has also made use of the NASA/IPAC Infrared Science Archive,
which is operated by the Jet Propulsion Laboratory, California Institute of
Technology, under contract with the National Aeronautics and Space Administration.
In addition, this research has made use of the SIMBAD database,
operated at CDS, Strasbourg, France 

\bibliographystyle{mn2e}
\bibliography{mn-jour,uchii-fix,mc-fix}

\begin{thebibliography}{}

\bibitem[\protect\citeauthoryear{{Acker}, {Marcout} \& {Ochsenbein}}{{Acker}
  et~al.}{1996}]{acker96}
{Acker} A.,  {Marcout} J.,    {Ochsenbein} F.,  1996, {First Supplement to the
  Strasbourg - ESO catalogue of galactic planetary nebulae}.
Garching: European Southern Observatory

\bibitem[\protect\citeauthoryear{{Acker}, {Marcout}, {Ochsenbein}, {Stenholm}
  \& {Tylenda}}{{Acker} et~al.}{1992}]{acker92}
{Acker} A.,  {Marcout} J.,  {Ochsenbein} F.,  {Stenholm} B.,    {Tylenda} R.,
  1992, {Strasbourg - ESO catalogue of galactic planetary nebulae}.
Garching: European Southern Observatory

\bibitem[\protect\citeauthoryear{{Afflerbach}, {Churchwell}, {Acord}, {Hofner},
  {Kurtz} \& {Depree}}{{Afflerbach} et~al.}{1996}]{afflerbach96}
{Afflerbach} A.,  {Churchwell} E.,  {Acord} J.~M.,  {Hofner} P.,  {Kurtz} S.,
   {Depree} C.~G.,  1996, ApJS, 106, 423

\bibitem[\protect\citeauthoryear{{Arenou}, {Grenon} \& {Gomez}}{{Arenou}
  et~al.}{1992}]{Arenou92}
{Arenou} F.,  {Grenon} M.,    {Gomez} A.,  1992, A\&A, 258, 104

\bibitem[\protect\citeauthoryear{{Benjamin}, {Churchwell}, {Babler}
  et~al.,}{{Benjamin} et~al.}{2003}]{bobben03}
{Benjamin} R.~A.,  {Churchwell} E.,  {Babler} B.~L.,    et~al., 2003, PASP,
  115, 953

\bibitem[\protect\citeauthoryear{{Bock}, {Large} \& {Sadler}}{{Bock}
  et~al.}{1999}]{bock99}
{Bock} D.~C.-J.,  {Large} M.~I.,    {Sadler} E.~M.,  1999, AJ, 117, 1578

\bibitem[\protect\citeauthoryear{{Carey} S.~J.~{Mizuno}, {Kraemer}, Shenoy,
  {Noriega-Crespo}, {Price}, Paladini \& {Kuchar}}{{Carey}
  et~al.}{2007}]{carey07}
{Carey} S.~J.~{Mizuno} D.~R.,  {Kraemer} K.~E.,  Shenoy S.,  {Noriega-Crespo}
  A.,  {Price} S.~D.,  Paladini R.,    {Kuchar} T.~A.,  2007, {MIPSGAL v1.0
  Data Delivery Description Document}, {http://data.spitzer.caltech.edu}

\bibitem[\protect\citeauthoryear{{Carey}, {Noriega-Crespo}, {Mizuno}
  et~al.,}{{Carey} et~al.}{2009}]{carey09}
{Carey} S.~J.,  {Noriega-Crespo} A.,  {Mizuno} D.~R.,    et~al., 2009, PASP,
  121, 76

\bibitem[\protect\citeauthoryear{{Carey}, {Noriega-Crespo}, {Price}
  et~al.,}{{Carey} et~al.}{2005}]{carey05}
{Carey} S.~J.,  {Noriega-Crespo} A.,  {Price} S.~D.,    et~al., 2005, in
  Bulletin of the American Astronomical Society Vol.~37, {MIPSGAL: A Survey of
  the Inner Galactic Plane at 24 and 70 microns, Survey Strategy and Early
  Results}.
pp 1--16

\bibitem[\protect\citeauthoryear{{Caswell} \& {Haynes}}{{Caswell} \&
  {Haynes}}{1987}]{caswell87}
{Caswell} J.~L.,  {Haynes} R.~F.,  1987, A\&A, 171, 261

\bibitem[\protect\citeauthoryear{{Chen}, {Vergely}, {Valette} \&
  {Carraro}}{{Chen} et~al.}{1998}]{Chen98}
{Chen} B.,  {Vergely} J.~L.,  {Valette} B.,    {Carraro} G.,  1998, A\&A, 336,
  137

\bibitem[\protect\citeauthoryear{{Churchwell}, {Watson}, {Povich}
  et~al.,}{{Churchwell} et~al.}{2007}]{churchwell07}
{Churchwell} E.,  {Watson} D.~F.,  {Povich} M.~S.,    et~al., 2007, {The
  Bubbling Galactic Disk II: the Inner 20 Degrees},
  {http://www.astro.wisc.edu/glimpse/BubblesII\_V2.pdf}

\bibitem[\protect\citeauthoryear{{Cohen} \& {Green}}{{Cohen} \&
  {Green}}{2001}]{cohen01}
{Cohen} M.,  {Green} A.~J.,  2001, MNRAS, 325, 531

\bibitem[\protect\citeauthoryear{{Cohen}, {Green}, {Meade} et~al.,}{{Cohen}
  et~al.}{2007}]{cohen07a}
{Cohen} M.,  {Green} A.~J.,  {Meade} M.~R.,    et~al., 2007, MNRAS, 374, 979

\bibitem[\protect\citeauthoryear{{Cohen}, {Parker}, {Green} et~al.,}{{Cohen}
  et~al.}{2007}]{cohen07}
{Cohen} M.,  {Parker} Q.~A.,  {Green} A.~J.,    et~al., 2007, ApJ, 669, 343

\bibitem[\protect\citeauthoryear{{Cohen}, {Walker} \& {Witteborn}}{{Cohen}
  et~al.}{1992}]{cohen92}
{Cohen} M.,  {Walker} R.~G.,    {Witteborn} F.~C.,  1992, AJ, 104, 2030

\bibitem[\protect\citeauthoryear{{Comeron} \& {Torra}}{{Comeron} \&
  {Torra}}{1996}]{comeron96}
{Comeron} F.,  {Torra} J.,  1996, A\&A, 314, 776

\bibitem[\protect\citeauthoryear{{Condon}}{{Condon}}{2007}]{condon07}
{Condon} J.~J.,  2007, Technical report, {Essential Radio Astornomy},
  \verb+http://www.cv.nrao.edu/course/astr534+.
National Radio Astronomy Observatory

\bibitem[\protect\citeauthoryear{{Fazio}, {Hora}, {Allen} et~al.,}{{Fazio}
  et~al.}{2004}]{fazio04}
{Fazio} G.~G.,  {Hora} J.~L.,  {Allen} L.~E.,    et~al., 2004, ApJS, 154, 10

\bibitem[\protect\citeauthoryear{{Gaume}, {Goss}, {Dickel}, {Wilson} \&
  {Johnston}}{{Gaume} et~al.}{1995}]{gaume95}
{Gaume} R.~A.,  {Goss} W.~M.,  {Dickel} H.~R.,  {Wilson} T.~L.,    {Johnston}
  K.~J.,  1995, ApJ, 438, 776

\bibitem[\protect\citeauthoryear{{Gibb} \& {Hoare}}{{Gibb} \&
  {Hoare}}{2007}]{gibb07}
{Gibb} A.~G.,  {Hoare} M.~G.,  2007, MNRAS, 380, 246

\bibitem[\protect\citeauthoryear{{Henning}, {Lapinov}, {Schreyer}, {Stecklum}
  \& {Zinchenko}}{{Henning} et~al.}{2000}]{henning00}
{Henning} T.,  {Lapinov} A.,  {Schreyer} K.,  {Stecklum} B.,    {Zinchenko} I.,
   2000, A\&A, 364, 613

\bibitem[\protect\citeauthoryear{{Hoare}}{{Hoare}}{2005}]{hoare05}
{Hoare} M.~G.,  2005, Astrophys. Space. Sci., 295, 203

\bibitem[\protect\citeauthoryear{{Hofner} \& {Churchwell}}{{Hofner} \&
  {Churchwell}}{1996}]{hofner96}
{Hofner} P.,  {Churchwell} E.,  1996, A\&AS, 120, 283

\bibitem[\protect\citeauthoryear{{Ignace} \& {Churchwell}}{{Ignace} \&
  {Churchwell}}{2004}]{ignace04}
{Ignace} R.,  {Churchwell} E.,  2004, ApJ, 610, 351

\bibitem[\protect\citeauthoryear{{Joshi}}{{Joshi}}{2005}]{Joshi05}
{Joshi} Y.~C.,  2005, MNRAS, 362, 1259

\bibitem[\protect\citeauthoryear{{Keto}}{{Keto}}{2003}]{keto03}
{Keto} E.,  2003, ApJ, 599, 1196

\bibitem[\protect\citeauthoryear{{Kurtz}}{{Kurtz}}{2002}]{kurtz02}
{Kurtz} S.,  2002, in {Crowther} P.,  ed., Hot Star Workshop III: The Earliest
  Phases of Massive Star Birth Vol.~267 of Astronomical Society of the Pacific
  Conference Series, {Ultracompact HII Regions}.
p.~81

\bibitem[\protect\citeauthoryear{{Kurtz}}{{Kurtz}}{2005}]{kurtz05}
{Kurtz} S.,  2005, in {Cesaroni} R.,  {Felli} M.,  {Churchwell} E.,
  {Walmsley} M.,  eds, Massive Star Birth: A Crossroads of Astrophysics
  Vol.~227 of IAU Symposium, {Hypercompact HII regions}.
pp 111--119

\bibitem[\protect\citeauthoryear{{Kurtz}}{{Kurtz}}{2000}]{kurtz00}
{Kurtz} S.~E.,  2000, in {Arthur} S.~J.,  {Brickhouse} N.~S.,   {Franco} J.,
  eds, Revista Mexicana de Astronomia y Astrofisica Conference Series Vol.~9 of
  Revista Mexicana de Astronomia y Astrofisica Conference Series, {Ultracompact
  H II Regions: New Challenges}.
pp 169--176

\bibitem[\protect\citeauthoryear{{Kurtz}, {Watson}, {Hofner} \& {Otte}}{{Kurtz}
  et~al.}{1999}]{kurtz99}
{Kurtz} S.~E.,  {Watson} A.~M.,  {Hofner} P.,    {Otte} B.,  1999, ApJ, 514,
  232

\bibitem[\protect\citeauthoryear{{Lizano}}{{Lizano}}{2008}]{lizano08}
{Lizano} S.,  2008, in {Beuther} H.,  {Linz} H.,   {Henning} T.,  eds, Massive
  Star Formation: Observations Confront Theory Vol.~387 of Astronomical Society
  of the Pacific Conference Series, {Hypercompact HII Regions}.
p.~232

\bibitem[\protect\citeauthoryear{{Marsh}}{{Marsh}}{1975}]{marsh75}
{Marsh} K.~A.,  1975, ApJ, 201, 190

\bibitem[\protect\citeauthoryear{{Marsh}}{{Marsh}}{1976}]{marsh76}
{Marsh} K.~A.,  1976, ApJ, 203, 552

\bibitem[\protect\citeauthoryear{{Massardi}, {Ekers}, {Murphy}
  et~al.,}{{Massardi} et~al.}{2008}]{massardi08}
{Massardi} M.,  {Ekers} R.~D.,  {Murphy} T.,    et~al., 2008, MNRAS, 384, 775

\bibitem[\protect\citeauthoryear{{Mauch}, {Murphy}, {Buttery}, {Curran},
  {Hunstead}, {Piestrzynski}, {Robertson} \& {Sadler}}{{Mauch}
  et~al.}{2003}]{mauch03}
{Mauch} T.,  {Murphy} T.,  {Buttery} H.~J.,  {Curran} J.,  {Hunstead} R.~W.,
  {Piestrzynski} B.,  {Robertson} J.~G.,    {Sadler} E.~M.,  2003, MNRAS, 342,
  1117

\bibitem[\protect\citeauthoryear{{McClure-Griffiths} \&
  {Dickey}}{{McClure-Griffiths} \& {Dickey}}{2007}]{mcclure-griffiths07}
{McClure-Griffiths} N.~M.,  {Dickey} J.~M.,  2007, ApJ, 671, 427

\bibitem[\protect\citeauthoryear{{Mezger} \& {Henderson}}{{Mezger} \&
  {Henderson}}{1967}]{mezger67}
{Mezger} P.~G.,  {Henderson} A.~P.,  1967, ApJ, 147, 471

\bibitem[\protect\citeauthoryear{{Mills}}{{Mills}}{1981}]{mills81}
{Mills} B.~Y.,  1981, Proceedings of the Astronomical Society of Australia, 4,
  156

\bibitem[\protect\citeauthoryear{{Murphy}, {Mauch}, {Green}, {Hunstead},
  {Piestrzynska}, {Kels} \& {Sztajer}}{{Murphy} et~al.}{2007}]{murphy07}
{Murphy} T.,  {Mauch} T.,  {Green} A.,  {Hunstead} R.~W.,  {Piestrzynska} B.,
  {Kels} A.~P.,    {Sztajer} P.,  2007, MNRAS, 382, 382

\bibitem[\protect\citeauthoryear{{Murphy}, {Sadler}, {Ekers} et~al.,}{{Murphy}
  et~al.}{2010}]{murphy10}
{Murphy} T.,  {Sadler} E.~M.,  {Ekers} R.~D.,    et~al., 2010, MNRAS, in press,
  (astro-ph/0911.0002)

\bibitem[\protect\citeauthoryear{{Olnon}}{{Olnon}}{1975}]{olnon75}
{Olnon} F.~M.,  1975, A\&A, 39, 217

\bibitem[\protect\citeauthoryear{{Panagia} \& {Felli}}{{Panagia} \&
  {Felli}}{1975}]{panagia75}
{Panagia} N.,  {Felli} M.,  1975, A\&A, 39, 1

\bibitem[\protect\citeauthoryear{{Parker}, {Phillipps}, {Pierce}
  et~al.,}{{Parker} et~al.}{2005}]{Parker05}
{Parker} Q.~A.,  {Phillipps} S.,  {Pierce} M.~J.,    et~al., 2005, MNRAS, 362,
  689

\bibitem[\protect\citeauthoryear{{Price}, {Egan}, {Carey}, {Mizuno} \&
  {Kuchar}}{{Price} et~al.}{2001}]{Price01}
{Price} S.~D.,  {Egan} M.~P.,  {Carey} S.~J.,  {Mizuno} D.~R.,    {Kuchar}
  T.~A.,  2001, AJ, 121, 2819

\bibitem[\protect\citeauthoryear{{Robertson}}{{Robertson}}{1991}]{robertson91}
{Robertson} J.~G.,  1991, Australian Journal of Physics, 44, 729

\bibitem[\protect\citeauthoryear{{Roche} \& {Aitken}}{{Roche} \&
  {Aitken}}{1984}]{RA84}
{Roche} P.~F.,  {Aitken} D.~K.,  1984, MNRAS, 208, 481

\bibitem[\protect\citeauthoryear{{Sabbatini}, {Cavaliere}, {dall'Oglio}
  et~al.,}{{Sabbatini} et~al.}{2005}]{sabbatini05}
{Sabbatini} L.,  {Cavaliere} F.,  {dall'Oglio} G.,    et~al., 2005, A\&A, 439,
  595

\bibitem[\protect\citeauthoryear{{Sewilo}, {Churchwell}, {Kurtz}, {Goss} \&
  {Hofner}}{{Sewilo} et~al.}{2004}]{sewilo04}
{Sewilo} M.,  {Churchwell} E.,  {Kurtz} S.,  {Goss} W.~M.,    {Hofner} P.,
  2004, ApJ, 605, 285

\bibitem[\protect\citeauthoryear{{Wood} \& {Churchwell}}{{Wood} \&
  {Churchwell}}{1989}]{wood89a}
{Wood} D.~O.~S.,  {Churchwell} E.,  1989, ApJS, 69, 831

\bibitem[\protect\citeauthoryear{{Wright} \& {Barlow}}{{Wright} \&
  {Barlow}}{1975}]{wright75}
{Wright} A.~E.,  {Barlow} M.~J.,  1975, MNRAS, 170, 41

\end{thebibliography}

\label{lastpage}

\end{document}